\documentclass[twocolumn,twocolumnappendix]{aastex631}
\received{2023 September 27}
\revised{2023 December 21}
\accepted{2024 January 09}  % merci à Pauline Jaricot :)!

\submitjournal{ApJ}

\shorttitle{Accretion Lines from TWA 27B}  %
\shortauthors{Marleau et al.}
\usepackage{etoolbox}

\makeatletter

\patchcmd{\NAT@citex}
  {\@citea\NAT@hyper@{%
     \NAT@nmfmt{\NAT@nm}%
     \hyper@natlinkbreak{\NAT@aysep\NAT@spacechar}{\@citeb\@extra@b@citeb}%
     \NAT@date}}
  {\@citea\NAT@nmfmt{\NAT@nm}%
   \NAT@aysep\NAT@spacechar\NAT@hyper@{\NAT@date}}{}{}

\patchcmd{\NAT@citex}
  {\@citea\NAT@hyper@{%
     \NAT@nmfmt{\NAT@nm}%
     \hyper@natlinkbreak{\NAT@spacechar\NAT@@open\if*#1*\else#1\NAT@spacechar\fi}%
       {\@citeb\@extra@b@citeb}%
     \NAT@date}}
  {\@citea\NAT@nmfmt{\NAT@nm}%
   \NAT@spacechar\NAT@@open\if*#1*\else#1\NAT@spacechar\fi\NAT@hyper@{\NAT@date}}
  {}{}

\makeatother
\def\mum{\ensuremath{\upmu\mathrm{m}}\xspace}

\usepackage{accsupp}

\makeatletter
\let\jnl@style=\rm
\def\ref@jnl#1{{\jnl@style#1}}

\def\aj{\ref@jnl{AJ}}                   %
\def\actaa{\ref@jnl{Acta Astron.}}      %
\def\araa{\ref@jnl{ARA\&A}}             %
\def\apj{\ref@jnl{ApJ}}                 %
\def\apjl{\ref@jnl{ApJ}}                %
\def\apjs{\ref@jnl{ApJS}}               %
\def\ao{\ref@jnl{Appl.~Opt.}}           %
\def\apss{\ref@jnl{Ap\&SS}}             %
\def\aap{\ref@jnl{A\&A}}                %
\def\aapr{\ref@jnl{A\&A~Rev.}}          %
\def\aaps{\ref@jnl{A\&AS}}              %
\def\azh{\ref@jnl{AZh}}                 %
\def\baas{\ref@jnl{BAAS}}               %
\def\bac{\ref@jnl{Bull. astr. Inst. Czechosl.}}
\def\caa{\ref@jnl{Chinese Astron. Astrophys.}}
\def\cjaa{\ref@jnl{Chinese J. Astron. Astrophys.}}
\def\icarus{\ref@jnl{Icarus}}           %
\def\jcap{\ref@jnl{J. Cosmology Astropart. Phys.}}
\def\jrasc{\ref@jnl{JRASC}}             %
\def\memras{\ref@jnl{MmRAS}}            %
\def\mnras{\ref@jnl{MNRAS}}             %
\def\na{\ref@jnl{New A}}                %
\def\nar{\ref@jnl{New A Rev.}}          %
\def\pra{\ref@jnl{Phys.~Rev.~A}}        %
\def\prb{\ref@jnl{Phys.~Rev.~B}}        %
\def\prc{\ref@jnl{Phys.~Rev.~C}}        %
\def\prd{\ref@jnl{Phys.~Rev.~D}}        %
\def\pre{\ref@jnl{Phys.~Rev.~E}}        %
\def\prl{\ref@jnl{Phys.~Rev.~Lett.}}    %
\def\pasa{\ref@jnl{PASA}}               %
\def\pasp{\ref@jnl{PASP}}               %
\def\pasj{\ref@jnl{PASJ}}               %
\def\rmxaa{\ref@jnl{Rev. Mexicana Astron. Astrofis.}}%
\def\qjras{\ref@jnl{QJRAS}}             %
\def\skytel{\ref@jnl{S\&T}}             %
\def\solphys{\ref@jnl{Sol.~Phys.}}      %
\def\sovast{\ref@jnl{Soviet~Ast.}}      %
\def\ssr{\ref@jnl{Space~Sci.~Rev.}}     %
\def\zap{\ref@jnl{ZAp}}                 %
\def\nat{\ref@jnl{Nature}}              %
\def\iaucirc{\ref@jnl{IAU~Circ.}}       %
\def\aplett{\ref@jnl{Astrophys.~Lett.}} %
\def\apspr{\ref@jnl{Astrophys.~Space~Phys.~Res.}}
\def\bain{\ref@jnl{Bull.~Astron.~Inst.~Netherlands}} 
\def\fcp{\ref@jnl{Fund.~Cosmic~Phys.}}  %
\def\gca{\ref@jnl{Geochim.~Cosmochim.~Acta}}   %
\def\grl{\ref@jnl{Geophys.~Res.~Lett.}} %
\def\jcp{\ref@jnl{J.~Chem.~Phys.}}      %
\def\jgr{\ref@jnl{J.~Geophys.~Res.}}    %
\def\jqsrt{\ref@jnl{J.~Quant.~Spec.~Radiat.~Transf.}}
\def\memsai{\ref@jnl{Mem.~Soc.~Astron.~Italiana}}
\def\nphysa{\ref@jnl{Nucl.~Phys.~A}}   %
\def\physrep{\ref@jnl{Phys.~Rep.}}   %
\def\physscr{\ref@jnl{Phys.~Scr}}   %
\def\planss{\ref@jnl{Planet.~Space~Sci.}}   %
\def\procspie{\ref@jnl{Proc.~SPIE}}   %

\def\ptp{\ref@jnl{Prog.~Th.~Phys.}}   %
\def\natas{\ref@jnl{NatAs}}           %
\def\amjm{\ref@jnl{AmJM}}             %

\makeatother

\defcitealias{goli04}{G04}
\defcitealias{marl07}{M07}
\defcitealias{scvh}{SCvH}
\defcitealias{bl94}{BL94}
\defcitealias{mc14}{MC14}
\defcitealias{ensman94}{E94}
\defcitealias{lever81}{LP81}

\defcitealias{m16Schock}{Paper~I}
\defcitealias{m18Schock}{Paper~II}
\defcitealias{m22Schock}{Paper~III}

\defcitealias{boden13}{B13}
\defcitealias{emsen21a}{NGPPS}

\defcitealias{ab22}{AB22}

\usepackage{xspace}

\usepackage{amsmath,amssymb,listings,upgreek,nicefrac}
\usepackage{newtxtext,newtxmath}
\usepackage{grffile}
\usepackage{xcolor,xspace}
\usepackage[normalem]{ulem}  %

\usepackage[T1]{fontenc}
\usepackage{textcomp}     %

\newcounter{numKommG}
\newcommand{\neuI}[1]{{\leavevmode{\boldmath\bfseries#1}}}
\newcommand{\neuII}[1]{{\leavevmode{\boldmath\bfseries#1}}}

\renewcommand{\neuI}[1]{{\leavevmode#1}}    
\renewcommand{\neuII}[1]{{\leavevmode#1}}

\newcommand{\neuIII}[1]{{\leavevmode\bfseries\neuII{#1}}}   %
\def\e{\textrm{e}}                                 %
\def\MJ{\ensuremath{M_{\textrm{J}}}\xspace}        %
\def\RJ{\ensuremath{R_{\textrm{J}}}\xspace}        %
\def\LSonne{\ensuremath{L_\odot}\xspace}           %

\def\Ha{\ensuremath{\mathrm{H}\,\alpha}\xspace}             %
\def\Paa{\ensuremath{\mathrm{Pa}\,\alpha}\xspace}           %
\def\Pab{\ensuremath{\mathrm{Pa}\,\beta}\xspace}            %
\def\Pag{\ensuremath{\mathrm{Pa}\,\gamma}\xspace}           %
\def\Pad{\ensuremath{\mathrm{Pa}\,\delta}\xspace}           %
\def\Bra{\ensuremath{\mathrm{Br}\,\alpha}\xspace}           %
\def\Brb{\ensuremath{\mathrm{Br}\,\beta}\xspace}            %
\def\Brg{\ensuremath{\mathrm{Br}\,\gamma}\xspace}           %
\def\Brd{\ensuremath{\mathrm{Br}\,\delta}\xspace}           %
\def\Bre{\ensuremath{\mathrm{Br}\,\epsilon}\xspace}         %
\def\Brz{\ensuremath{\mathrm{Br}}\,10\xspace}               %
\def\Pfa{\ensuremath{\mathrm{Pf}\,\alpha}\xspace}           %
\def\Pfb{\ensuremath{\mathrm{Pf}\,\beta}\xspace}            %
\def\Pfg{\ensuremath{\mathrm{Pf}\,\gamma}\xspace}           %
\def\Pfd{\ensuremath{\mathrm{Pf}\,\delta}\xspace}           %
\def\Pfe{\ensuremath{\mathrm{Pf}\,\epsilon}\xspace}         %
\def\HeI{\ensuremath{\mathrm{He}}\,\textsc{i}\xspace}       %
\def\HeIt{\HeI}                     %
\def\HeItVak{\HeI~\ensuremath{\lambda}10833\xspace}         %
\def\PDSbc{PDS\,70\,b and~c\xspace}                       %
\def\Dlrmb{Delorme\,1\,(AB)\,b\xspace}                      %

\def\MPkt{\ensuremath{\dot{M}}\xspace}                               %
\def\MPktA{\ensuremath{\dot{M}_{\textrm{A}}}\xspace}                  %
\def\MPktB{\ensuremath{\dot{M}_{\textrm{B}}}\xspace}                  %
\def\MP{\ensuremath{M_{\textrm{p}}}\xspace}        %
\def\RP{\ensuremath{R_{\textrm{p}}}\xspace}        %
\newcommand{\Lacc}{\ensuremath{{L_{\textrm{acc}}}}\xspace}  %
\def\LAkk{\Lacc}
\def\FHa{\ensuremath{F_{\textrm{H}\,\alpha}}\xspace}             %
\def\FPab{\ensuremath{F_{\textrm{Pa}\,\beta}}\xspace}    %
\newcommand{\Fldichte}{\ensuremath{\mathcal{F}}\xspace} %
\def\FKont{\ensuremath{\Fldichte_{\textrm{cont}}}\xspace}       %
\def\sFKont{\ensuremath{\sigma_{\Fldichte_{\textrm{cont}}}}\xspace} %
\def\FLinienspitze{\ensuremath{\Fldichte_{0}}\xspace}           %
\def\FLinie{\ensuremath{F_{\textrm{line}}}\xspace}                %
\def\FLiniemitKont{\ensuremath{\FLinie^\star}\xspace}      %
\def\LLinie{\ensuremath{L_{\textrm{line}}}\xspace}                      %
\newcommand{\RH}{\ensuremath{{R_{\textrm{Hill}}}}\xspace}       %

\newcommand{\Teff}{\ensuremath{T_{\textrm{eff}}}\xspace}         %

\newcommand{\vFfinfty}{\ensuremath{{\varv_{\textrm{ff},\,\infty}}}\xspace}

\def\Dv{\ensuremath{\Delta \varv}\xspace}

\def\DvKont{\ensuremath{\Delta \varv_{\mathrm{cont}}}\xspace}
\def\DvInst{\ensuremath{\Delta \varv_{\mathrm{inst}}}\xspace}
\def\DvMask{\ensuremath{\Delta \varv_{\mathrm{mask}}}\xspace}
\def\Dvint{\ensuremath{\Delta \varv_{\mathrm{intrsc}}}\xspace}

\def\chiqkrit{\ensuremath{\chi^2_{\textrm{crit}}}\xspace}
\def\Sigexpq{\ensuremath{\mathfrak{S}}\xspace}  %
\def\FOberg{\ensuremath{F_{\textrm{line}}^{\textrm{upp}}}\xspace}
\def\NPix{\ensuremath{N_{\textrm{pix}}}\xspace}
\def\DlmbdPix{\ensuremath{\Delta\lambda_\textrm{disp}}\xspace}

\def\RDv{\ensuremath{\mathcal{R}_{\Delta \varv}}\xspace}

\def\kms{\ensuremath{\textrm{km}\,\textrm{s}^{-1}}\xspace}    %
\def\MPktEJ{\ensuremath{\MJ\,\textrm{yr}^{-1}}\xspace}        %
\def\FEcgs{\ensuremath{\textrm{erg}\,\textrm{s}^{-1}\,\textrm{cm}^{-2}}\xspace}

\def\FdEcgs{\ensuremath{\textrm{erg}\,\textrm{s}^{-1}\,\textrm{cm}^{-2}\,\textrm{\AA}^{-1}}\xspace}

\renewcommand{\arraystretch}{1.4}

\usepackage{booktabs}  %

\def\twa{TWA~27A\xspace}
\def\twb{TWA~27B\xspace}

\begin{document}

\title{Revisiting the Helium and Hydrogen Accretion Indicators at TWA 27B:
Weak %
Mass Flow at Near-Freefall Velocity}

\author[0000-0002-2919-7500]{Gabriel-Dominique Marleau}
\affiliation{
Fakult\"at f\"ur Physik,
Universit\"at Duisburg-Essen,
Lotharstra\ss{}e 1,
47057 Duisburg, Germany; \href{mailto:gabriel.marleau@uni-tuebingen.de,gabriel.marleau@uni-due.de}{gabriel.marleau@uni-\{due,tuebingen\}.de}
}
\affiliation{%
Institut f\"ur Astronomie und Astrophysik,
Universit\"at T\"ubingen,
Auf der Morgenstelle 10,
72076 T\"ubingen, Germany
}
\affiliation{%
Physikalisches Institut,
Universit\"{a}t Bern,
Gesellschaftsstr.~6,
3012 Bern, Switzerland}
\affiliation{%
Max-Planck-Institut f\"ur Astronomie,
K\"onigstuhl 17,
69117 Heidelberg, Germany
}

\author[0000-0003-0568-9225]{Yuhiko Aoyama}
\affiliation{
Kavli Institute for Astronomy and Astrophysics, Peking University, Beijing 100084, People’s Republic of China
}

\author[0000-0002-3053-3575]{Jun Hashimoto}
\affiliation{
Astrobiology Center, National Institutes of Natural Sciences, 2-21-1 Osawa, Mitaka, Tokyo 181-8588, Japan
}
\affiliation{
Subaru Telescope, National Astronomical Observatory of Japan, Mitaka, Tokyo 181-8588, Japan
}
\affiliation{
Department of Astronomy, School of Science, Graduate University for Advanced Studies (SOKENDAI), Mitaka, Tokyo 181-8588, Japan
}

\author[0000-0003-2969-6040]{Yifan Zhou}
\affiliation{
Department of Astronomy,
University of Virginia,
530 McCormick Rd,
Charlottesville, VA 22904,
USA
}

\begin{abstract}
\twb (2M1207b) is the first directly-imaged planetary-mass ($\MP\approx5~\MJ$) companion \citep{chauvin04} and was observed at 0.9--5.3~\mum with JWST/NIRSpec \citep{luhman23c}.
To understand the accretion properties of \twb, we search for
continuum-subtracted   %
near-infrared helium and hydrogen emission lines and measure their widths and luminosities.
We detect the \HeIt  %
triplet at $4.3\sigma$ and
all Paschen-series lines covered by NIRSpec (\Paa, \Pab, \Pag, \Pad)
at 4--$5\sigma$.
The three brightest Brackett-series lines (\Bra, \Brb, \Brg) as well as \Pfg and \Pfd 
are tentative detections at 2--3$\sigma$.
We provide upper limits on the other hydrogen lines, \neuI{including on \Ha through Hubble Space Telescope archival data}.
Three lines can be reliably deconvolved to reveal an intrinsic width $\Dvint=(67\pm9)~\kms$, which is 60\,\%\ of the surface freefall velocity.
The line luminosities
\neuI{seem significantly too high to be}
due to chromospheric activity.
\neuI{Converting} line luminosities \neuI{to} an accretion rate \neuI{yields} $\MPkt\approx5\times10^{-9}~\MPktEJ$ when using scaling relationships for planetary masses,
and $\MPkt\approx0.1\times10^{-9}~\MPktEJ$ with extrapolated stellar scalings.
Several of these lines represent first detections at an accretor of such low mass. The weak accretion rate implies that formation is likely over.
This analysis shows that JWST can be used to measure \neuI{low} line-emitting mass accretion rates \neuI{onto} planetary-mass objects, motivates deeper searches for the mass reservoir feeding \twb,
and hints that other young directly-imaged objects might---hitherto unbeknownst---also be accreting.
\end{abstract}


\section{Introduction}
 \label{sec:intro}

\neuI{%
While several hundreds of substellar objects display convincing evidence of ongoing accretion \citep{betti23}, only a few have masses below $\MP\approx10~\MJ$ and have been studied extensively in the last few years:
\PDSbc \citep{wagner18,Haffert+2019}
and \Dlrmb \citep{eriksson20,betti22b,betti22c,ringqvist23}.
It is therefore capital to enlarge this population to study the dependence of the accretion rate on object mass and age, with the prospect of better understanding the differences in formation processes from stars down to planets.
}

\neuI{%
Brown dwarfs and gas giants likely gain mass through different but not necessarily mutually exclusive physical mechanisms: 
magnetospheric accretion from a local gas reservoir \citep[e.g.,][]{calvetgull98,hartmann16,thanathibodee19,hasegawa24}
or
large-scale direct accretion onto the surface of the object and its circumplanetary (or ``circum-substellar-object disc'') disc (CPD) \citep[e.g.,][]{tanigawa12,aoyama18,m22Schock}.
These modes of accretion are expected to produce line emission, strongest in the lines of neutral hydrogen.
The mass of the accretor is a key determinant of the physical conditions at the accretion shock on its surface or on the CPD. In turn, these conditions lead to predictions allowing one to distinguish the mechanisms. This is starting to be leveraged \citep{demars23}, but line-resolved ($R\gtrsim15,000$) observations will be required to take advantage of the full diagnostic potential \citep{maea21}.
}

\neuI{%
If only an integrated line luminosity is available, as is often the case, care must be taken to use an appropriate scaling relationship to estimate the accretion rate from the line luminosity \citep{betti23}.
Blindly extrapolating stellar relationships does not seem justified and in fact likely underestimates accretion rates systematically, by up to a few orders of magnitude \citep{AMIM21L,ma22}.
Nevertheless, both for spectroscopically resolved and non-resolved (low-resolution or photometric) observations, sensitivity to low line fluxes is critical to avoid being biased towards high fluxes and thus accretion rates; only non-biased observations allow meaningful studies of the scatter in the $M(\MPkt)$ correlation and its dependence on physical parameters such as age \citep{betti23}.
}

\neuI{%
In a wider context, the James Webb Space Telescope (JWST; \citealp{gardner23}) has a tremendous potential to help answer outstanding questions in planetary accretion and formation.
The advantages of JWST are manifold.
Firstly, JWST can access bright hydrogen $\alpha$-transitions, such as \Paa and \Bra (respectively at 1.875~\mum and 4.051~\mum), which are challenging to observe from the ground due to Earth's atmospheric absorption. Traditionally, infrared (IR) observations have played a crucial role as a powerful probe in star-forming regions with high extinction, which are not amenable to UV and visible observations.
Secondly, medium-resolution spectroscopy ($R\sim2,000$) using an integral field unit (IFU) with wide wavelength coverage enables the direct detection of multiple IR emission lines in substellar companions. Near-simultaneous observations are devoid of short-term variability (e.g., GQ Lup~B; \citealp{demars23}), enabling the acquisition of reliable results from the analysis of multiple lines.
Lastly, JWST provides diffraction-limited observations, resulting in unprecedented sensitivity. This is a particular advantage for observing isolated substellar objects that are generally challenging to observe with ground-based adaptive optics (AO) due to their intrinsic faintness. JWST will characterize more accreting free-floating planets. For reference, the JWST sensitivity at NIR wavelengths is approximately 100~times better than from the ground (see the ``Historical Sensitivity Estimates'' on the JWST website).  %
These observations are crucial for understanding very faint lines from substellar objects in a quiescent accretion phase \citep[e.g.,][]{brittain20} or with very weak accretion close to the end of their formation, which might occur earlier than thought (e.g., \citealp{wagner23}).%
}

With its mass of $\MP\approx5~\MJ$, 
\twb, also known as 2M1207b, is the first planetary-mass directly-imaged object. It was discovered by \citet{chauvin04}, two years after identification of the primary as a probably accreting member of the TW Hydr\ae{} Association (TWA) by \citet{gizis02}, and followed four years later by the iconic
HR~8799 system \citep{marois08,marois10}.
However, \twb\ \neuI{has a mass ratio of only $q\approx0.2$ relative to} the primary \twa, \neuI{which suggests the} system might have formed \neuI{not in a planetary but rather a stellar way \citep{lodato05,mohanty13,reggiani16,bowler20}}.

\citet{luhman23c} \neuI{analysed the GTO~1270 (PI: S.~Birkmann) JWST/NIRSpec \citep{jakobsen22} integral-ﬁeld unit \citep{boeker22} data on} \twb. Thanks to clear hydrogen-line emission, they showed that this object is accreting. They focused on the photospheric emission and measured the flux contained in one hydrogen transition, \Pab.
Here, we search systematically for hydrogen lines and report the continuum-corrected line fluxes. We then use scaling relationships between line and accretion luminosity to estimate the accretion rate, \neuI{and argue robustly that the line fluxes are not dominated by chromospheric activity}.

\section{Data and Methods}
 \label{sec:datmeth}

\subsection{Data sources}

The JWST data were presented in \citet{luhman23c} and were obtained with the NIRSpec instrument. They consist of one-hour exposures starting on
2023 February 07 at 17:52:28.779 with the G395H/F290LP
\neuI{high-resolution-}grating--filter combination
(0.97--1.89~\mum; \neuI{after this overview the filter names will be dropped for conciseness}),
at 18:32:50.598 with G235H/F170LP (1.66--3.17~\mum), and
at 19:11:59.463 with \neuI{G140H/F100LP (2.87--5.27~\mum)}.
The spectrum for each half-grating and the errorbars were kindly shared by K.~Luhman.
Some of the flux errorbars were spuriously high and we replaced them with the median of the well-behaved errors.
In any case, the dominant source of noise is the continuum noise, as shown below.

\neuI{%
In addition to JWST data, we provide an \Ha photometric point from our analysis of Hubble Space Telescope (HST) archival data.
HST observed the TWA~27 system with WFC3/UVIS on 2011 March 28 (PID: 12225, PI: A.~Reiners) and we show the F656N image in Appendix~\ref{sec:HSTHa}.
The HST image does not yield a significant detection of \twb\ \neuII{(only a tentative one)} and provides an upper limit on its \Ha emission. We will use this in Section~\ref{sec:Lacc}.%
}

\subsection{Fitting approach}

We searched \neuI{the JWST data} for a signal at the metastable neutral helium triplet \HeItVak as well as all hydrogen lines covered by the detectors in the Paschen-transition series (electron final energy level $n_f=3$), which is $n_i=4$--7 (\Paa--\Pad); all lines up to an initial level $n_i=10$ (\Brz) in the Brackett series ($n_f=4$); and all lines up to $n_i=10$ (\Pfe) in the Pfund series ($n_f=5$), \neuI{but without} \Pfa ($n_i=6$) \neuI{since it is off the red edge of NIRSpec} with $\lambda_0=7.46~\upmu$m.
\neuI{An overview of the lines is in Table~\ref{tab:main}.}
From the Humphreys series ($n_f=6$), only the transitions $n_i=10$--12 or~13 (5.128--4.170~or 4.376~\mum, respectively; $n_i=13$ is on the detector edge) are covered by NIRSpec and visual inspection did not reveal any significant peak at the respective locations. Therefore, we do not analyse this series.

\begin{figure*}
 \centering
 \includegraphics[width=0.43\textwidth]{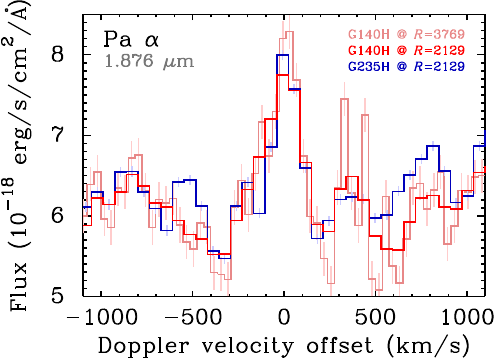}~~\includegraphics[width=0.515\textwidth]{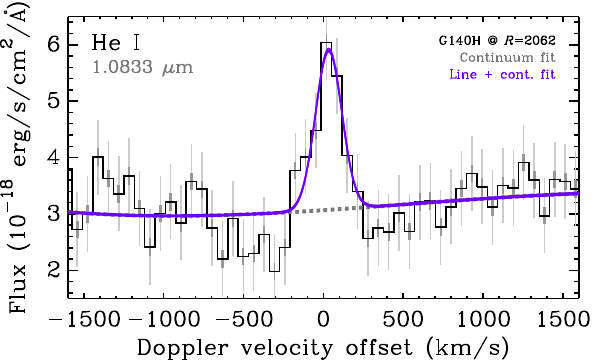}
\caption{%
\textit{Left:}
The \Paa line through the red half of the G140H (red) and the blue half of the G235H (blue) gratings at full resolution (pale red and bright blue), and from G140H down-convolved and re-interpolated to the resolution of G235H (bright red).
\textit{Right:}
\HeIt data (black) with errorbars (dark grey: only from each bin; pale grey: adding the continuum RMS in quadrature), cubic fit to the continuum (grey dashed line), and fit to the continuum and line (purple).
The continuum RMS noise is shown in Figure~\ref{fig:lineprofHeIt}.
}
\label{fig:Paa+HeIt}
\end{figure*}

In Figure~\ref{fig:Paa+HeIt}a, we show as an example the \Paa line, covered by G140H and G235H.
We show the data from each grating but also the data from G140H %
downgraded to the resolution of G235H.
For this, we convolved the G140H data with a Gaussian of width \Dv given by $\Dv^2=\Dv_2^2-\Dv_1^2$,
where $\Dv_1$ ($\Dv_2$) corresponds to the resolution of G140H (G235H) at \Paa, $R=3769$ ($R=2129$), as reported in Table~\ref{tab:main}.
As a reminder, the Gaussian function used for smoothing and as the line profile shape in the analysis
is $X=X_0\exp{[-0.5(\varv/\Dv)^2]}$, where $X$ is the filter height or the flux density \neuI{and $\varv$ is the Doppler velocity offset from the vacuum central wavelength}.
When compared at similar resolution, the flux levels are mostly consistent in absolute value.
A few features are seen in the down-resolved G140H data but not in G235H, suggesting that the errorbars are still somewhat estimated. However, the overall agreement is very good.

For each line, we fitted the local continuum by a cubic function and the line excess by a Gaussian profile, using  \texttt{gnuplot}'s built-in \texttt{fit} routine.
We took a three-step approach:
\begin{enumerate}
    \item 
We fitted the continuum in a range $\DvKont=2700~\kms$ on either side of the theoretical vacuum line centre \citep{Wiese+Fuhr2009}, masking out $|\varv|\equiv\DvMask=300~\kms$.
This value of \DvKont prevents the \HeIt and \Pag lines, separated by only about 2900~\kms, from interfering with each other, and \DvMask is a few times the expected line width for an $\MP\approx5~\MJ$, $\RP\approx(1$--$2)\RJ$ object \citep{luhman23c}, which is conservatively at most $\Delta \varv\approx\vFfinfty$ \citep{aoyama18}, where $\vFfinfty=\sqrt{2G\MP/\RP}\approx120~\kms$ is the free-fall velocity from infinity.
For two lines, the grating edge shortened slightly the actual wavelength \neuI{range} used for the continuum on the red side (\Bra at +1300~\kms and \Paa at +2200~\kms), compared to the nominal $\DvKont=2700~\kms$, but this is inconsequential.

An important outcome of this fitting is the root-mean square (RMS) deviation from the fitted continuum \neuI{(more properly, the \textit{standard deviation}, but we adopt the common usage)}. This continuum RMS is a source of noise because we do not attempt to identify and remove spectral features.

\item We fitted each continuum-subtracted line with a Gaussian function, again taking only the bin errorbars into account.

\item We initialised the fit parameters of the continuum and line excess to the values from the sequential fits and did a joint fit.
For the errorbars of each bin, we added in quadrature to the errorbar of each bin the RMS of the continuum.
The fits converged in around $\approx5$--20 iterations, typically around 10.

\end{enumerate}
An example outcome is shown in Figure~\ref{fig:Paa+HeIt}b.
We varied \DvKont and \DvMask and obtained essentially the same fit results for all the lines (not shown).
We note that the continuum RMS dominates the noise budget (see the pale and dark grey errorbars in Figure~\ref{fig:Paa+HeIt}b).

As a comparison, we also performed the fits using only the flux uncertainty in each bin as the errorbar. Some fitted lines were narrower than the instrumental broadening, but within only $1\sigma$, and the overall results were very similar.
Other variations  %
in the approach also led, if at all, to small differences of at most a few \kms in the line widths.
Thus the precise treatment of the errorbars does not matter in this case.

This fitting approach yields line fluxes, always meant as an excess above the fitted continuum (see Section~\ref{sec:fluxes}). The integrated luminosity of a given line is $\LLinie=4\pi d^2 \FLinie$, with $d=65.4$~pc the distance to \twb (see \citealp{luhman23c} and references therein) and $\FLinie=\sqrt{2\pi}\times\Dv\times\lambda_0/c\times \FLinienspitze$,
where $\Delta \varv$ is the fitted Doppler-shift velocity width,
\FLinienspitze is the flux-density peak value \textit{in excess of} the continuum,
$c$ is the speed of light, and $\lambda_0$ is the central vacuum wavelength of the transition. 

For simplicity and clarity, we do not attempt to correct the fluxes for possible extinction.
Interstellar extinction towards the TW Hydr\ae\ Association (TWA) and the TWA~27 system is negligible \citep{herczeg04,mohanty07}. Extinction by an edge-on disc around \twb has been suggested to reconcile the tension between the SED-derived \Teff and theoretical predictions (e.g., \citealp{mohanty07}). However, the analysis of \citet{luhman23c} suggests that using other atmospheric models instead might obviate the need for heavy extinction. Otherwise, somewhat ``tuned'' cloud properties could explain the spectrum and brightness of \twb \citep{skemer11}. The ultimate answer is not settled, and the line-emitting regions might be differently extincted than the atmosphere.
Thus keeping the fluxes ``as is'' therefore avoids introducing uncertainty.

\subsection{Detections and non-detections}
 \label{sec:detectnondetect}

In this work, we qualify a line as detected only if it has more than two bins at more than $3\sigma$. This is less stringent than the often-used $5\sigma$ criterion but appears justified given the low level of surprise, the modest impact of a detection, and the negligibility of the ``Look-Elsewhere Effect'', to use the considerations discussed by \citet{lyons13}. Unaccounted-for systematics will reduce the true significance of the detections but only modestly since the systematics seem small.

Other lines detected at 2--$3\sigma$ will be considered only tentative detections.
We report their fit parameters but do not further analyse them. Finally, for the other lines for which there is clearly no signal, we compile only upper limits on the flux and luminosity, and no fit parameters.
\neuI{%
We calculate upper limits by using an upper-tail one-sided test based on the $\chi^2$ distribution. Specifically, we solve for the minimum line flux which, broadened to the resolution at that wavelength, would lead to a $3\sigma$ deviation from a null excess:
\begin{subequations}
\begin{align}
    \FOberg &= \frac{\DvInst\lambda_0}{c}\times\sFKont\times\sqrt{\frac{2\pi\chiqkrit}{\Sigexpq}}\\
    \Sigexpq &\equiv \sum_{\mathrm{bins}} \exp^2\left[-\frac{1}{2}\left(\frac{\varv}{\DvInst}\right)^2\right],
\end{align}
\end{subequations}
where $\DvInst = c/R/(2\sqrt{2\ln 2})$  %
is the effective line width given by the} instrumental resolution\footnote{Obtained from the (pre-launch; dated 2016 August~30) ``Dispersion curves for the NIRSpec dispersers'' section of the JWST website.} \neuI{$R(\lambda_0)$,
\sFKont is the RMS of the fitted continuum
\neuI{(i.e., its standard deviation), and}
$\chiqkrit\approx33$ is the critical value of the $\chi^2$ distribution
that corresponds to $3\sigma$ (probability of 99.7\,\%) for $\nu\approx14-1$ degrees of freedom, since around 13--14~spectral bins are found within a (somewhat arbitrarily-chosen) range of $\pm3\DvInst$ around $\lambda_0$.
We
compute \Sigexpq for each line \neuII{(note the squared exponential)} and always find $\Sigexpq\approx 1.6$--1.7.  %
Thus
$\FOberg \approx 11.4 \times \sFKont \times
\DvInst\lambda_0/c%
$.}
A more detailed analysis could use the injection of fake planets and take the spatially varying sensitivity into account (e.g., \citealp{bonse23}).
However, since we will not analyse the non-detections in detail, we keep the simpler approach here.
\neuI{%
It is in principle more accurate than but comparable to the expression of \citet{betti22b},
$\FOberg=3\sqrt{\NPix}\times\sFKont\times\DlmbdPix=3$--4~$\textrm{\AA}\times\sFKont$,
where $\DlmbdPix=1.4$--2.8~\AA\,pixel$^{-1}$ (depending on the transition) is their spectral dispersion and $\NPix=(7~\textrm{\AA})/\DlmbdPix$,
or to the expression of \citet{alcal14} or \citet{gangi22},
$\FOberg=3\Delta\lambda\times\sFKont$ with $\Delta\lambda=1$--2~\AA{} their assumed line width.%
}

\section{Results}
 \label{sec:res}

\subsection{Line fluxes and widths}
 \label{sec:fluxes}

\begin{deluxetable*}{l c c c c c c c c c}  %
\tablecaption{
Helium-triplet and hydrogen lines covered by \citet{luhman23c}'s JWST spectrum of \twb.
\label{tab:main}}
\tablehead{%
 Line & $\lambda_0$ & $\FKont/10^{-18}$ & $\FLinie/10^{-17}$ & \LLinie & $R$ & $N_\sigma^F$ & $\varv_0$ & $\Delta \varv$ & \RDv \\
 & (\mum) & ($\FdEcgs$) & ($\FEcgs$) & ($10^{-9}~\LSonne$) & & & (\kms) & (\kms) &
}  %
\startdata 
 $\HeIt_{\textrm{1}}$ & 1.083 & $3.05 \pm 0.65$ & $2.16 \pm 0.50$ & $2.81 \pm 0.65$ & 2062 & 4.3 & $32\pm12$ & $ 84 \pm 13 $ & 1.4 \\ %
\hline
 $\Paa_{\textrm{1}}$ & 1.876 & $6.02 \pm 0.61$ & $2.41 \pm 0.67$ & $3.13 \pm 0.88$ & 3769 & 3.6 & $6\pm12$ & $ 70 \pm 13 $ & 2.1 \\ %
 $\Paa_{\textrm{2}}$ & 1.876 & $6.33 \pm 0.41$ & $1.44 \pm 0.33$ & $1.88 \pm 0.43$ & 2129 & 4.4 & $11\pm11$ & $ 50 \pm 12 $ & 0.8 \\ %
 $\Pab_{\textrm{1}}$ & 1.282 & $5.98 \pm 0.62$ & $2.72 \pm 0.61$ & $3.54 \pm 0.79$ & 2461 & 4.5 & $17\pm10$ & $ 91 \pm 11 $ & 1.8 \\ %
 $\Pag_{\textrm{1}}$ & 1.094 & $3.20 \pm 0.50$ & $1.57 \pm 0.31$ & $2.04 \pm 0.40$ & 2084 & 5.1 & $24\pm9$ & $ 67 \pm 9 $ & 1.1 \\ %
 $\Pad_{\textrm{1}}$ & 1.005 & $2.12 \pm 0.61$ & $1.58 \pm 0.40$ & $2.06 \pm 0.52$ & 1908 & 3.9 & $17\pm16$ & $ 78 \pm 16 $ & 1.2 \\ %
\hline
 $\Bra_{\textrm{3}}$ & 4.052 & $3.81 \pm 0.10$ & $0.56 \pm 0.24$ & $0.73 \pm 0.31$ & 2780 & 2.3 & $47\pm25$ & $ 70 \pm 26 $ & 1.5 \\ %
 $\Brb_{\textrm{2}}$ & 2.626 & $6.05 \pm 0.23$ & $0.73 \pm 0.23$ & $0.95 \pm 0.30$ & 3052 & 3.2 & $-16\pm14$ & $ 46 \pm 14 $ & 1.1 \\ %
 $\Brg_{\textrm{2}}$ & 2.166 & $8.75 \pm 0.22$ & $0.50 \pm 0.21$ & $0.65 \pm 0.27$ & 2478 & 2.4 & $5\pm22$ & $ 51 \pm 23 $ & 1.0 \\ %
 $\Brd_{\textrm{2}}$ & 1.945 & $6.69 \pm 0.32$ & $<1.32$ & $<1.72$ & 2213 & 0.7 & --- & --- & --- \\ %
 $\Bre_{\textrm{1}}$ & 1.818 & $6.80 \pm 0.62$ & $<1.47$ & $<1.92$ & 3629 & 0.2 & --- & --- & --- \\ %
 $\Bre_{\textrm{2}}$ & 1.818 & $6.55 \pm 0.35$ & $<1.48$ & $<1.92$ & 2061 & 0.4 & --- & --- & --- \\ %
 $\Brz_{\textrm{1}}$ & 1.737 & $7.90 \pm 0.56$ & $<1.36$ & $<1.76$ & 3438 & 0.5 & --- & --- & --- \\ %
 $\Brz_{\textrm{2}}$ & 1.737 & $7.85 \pm 0.36$ & $<1.52$ & $<1.97$ & 1965 & 2.2 & --- & --- & --- \\ %
\hline
 $\Pfb_{\textrm{3}}$ & 4.654 & $2.71 \pm 0.07$ & $<0.48$ & $<0.62$ & 3239 & 1.4 & --- & --- & --- \\ %
 $\Pfg_{\textrm{3}}$ & 3.740 & $4.80 \pm 0.08$ & $0.21 \pm 0.07$ & $0.27 \pm 0.09$ & 2549 & 3.0 & $-32\pm17$ & $ 27 \pm 52 $ & 0.6 \\ %
 $\Pfd_{\textrm{3}}$ & 3.297 & $5.25 \pm 0.10$ & $0.35 \pm 0.14$ & $0.45 \pm 0.18$ & 2231 & 2.6 & $-35\pm21$ & $ 49 \pm 24 $ & 0.9 \\ %
 $\Pfe_{\textrm{2}}$ & 3.039 & $5.25 \pm 0.20$ & $<0.81$ & $<1.05$ & 3610 & 1.0 & --- & --- & --- \\ %
 $\Pfe_{\textrm{3}}$ & 3.039 & $5.20 \pm 0.14$ & $<0.96$ & $<1.25$ & 2050 & 2.1 & --- & --- & --- \\ %
\enddata
\tablecomments{%
We consider all hydrogen lines covered by NIRSpec, but limited to an initial level $n_i\leqslant10$ in the Brackett ($n_f=4$) and Pfund ($n_f=5$) series.
The subscripts indicate the grating (1:~G140H, 2:~G235H, 3:~G395H), %
with \Paa, \Bre, \Brz, and \Pfe covered by two gratings.
\FKont is the fitted continuum level
at $\lambda_0$, with uncertainty \sFKont given by the RMS computed
over $\DvKont =\pm 2700$~\kms (see text).
\FLinie and \LLinie are the continuum-subtracted, line-integrated flux and luminosity.
$R$ is the spectral resolution at $\lambda_0$. %
$N_\sigma^F$ is the significance of the integrated line flux or the peak significance;
the peak flux density is $\Fldichte=\FKont+\FLinienspitze=\FKont+N_\sigma^F \sFKont$.
$\Delta \varv$ \neuI{and $\varv_0$ are} the fitted Doppler line width \neuI{and velocity offset of the peak}.
\RDv is the fitted line width relative to the instrumental broadening;
values below one suggest the continuum is overestimated.
For lines clearly not detected (see Section~\ref{sec:detectnondetect}),
\FLinie and \LLinie are $3\sigma$ upper limits (see text).%
}
\end{deluxetable*}

\begin{figure*}
 \centering
 \includegraphics[width=0.9\textwidth]{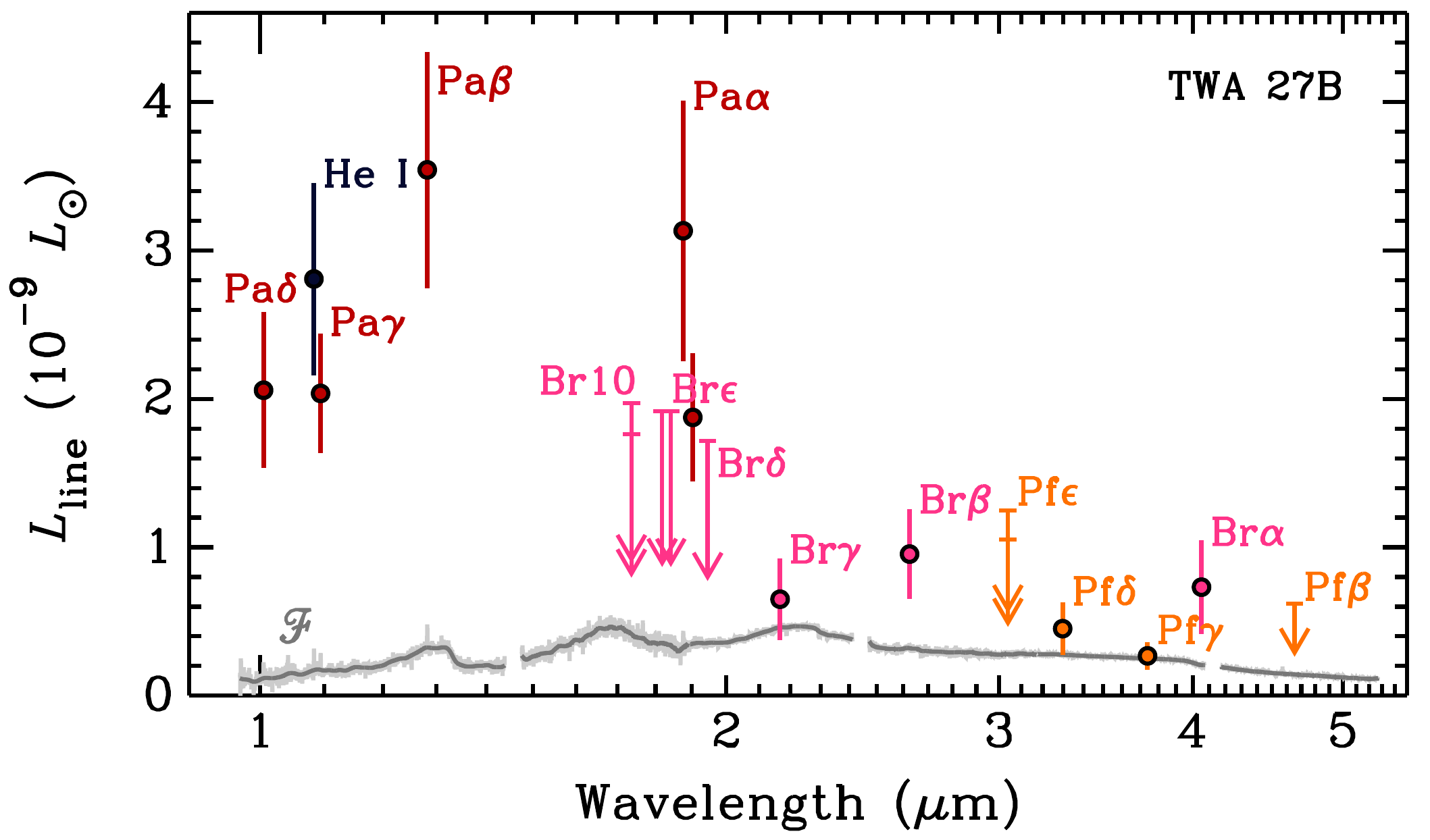}
\caption{%
Summary of the line luminosities in Table~\ref{tab:main}, coloured by series. Thick errorbars are for clear or tentative detections, and thin arrows for $3\sigma$ upper limits (see text).
$\Paa_2$ and \neuI{$\Bre_1$ are} shifted for clarity.
The $3\sigma$ upper limit on \Ha, $\LLinie<6.4\times10^{-9}~\LSonne$, is not shown.
The spectral density \Fldichte (at full resolution and smoothed; pale and dark grey, respectively) is shown against an arbitrary linear flux scale.
\neuI{The small gaps in the spectral coverage are detailed in \citet{luhman23c}.}
}
\label{fig:linesoverview}
\end{figure*}

We detect the neutral helium triplet at $4.3\sigma$ and
all Paschen-series lines covered by the observations (\Paa, \Pab, \Pag, \Pad)
at $>3.5$--$5\sigma$.
The three brightest Brackett-series lines (\Bra, \Brb, \Brg) as well as \Pfg and \Pfd are tentative detections at 2--3$\sigma$.
For the remaining hydrogen lines accessible to NIRSpec, we obtain upper limits.
All detections and non-detections are presented in Table~\ref{tab:main}, and illustrated in Figure~\ref{fig:linesoverview}, with line profiles in Appendix~\ref{sec:alllineprofiles}.
One of the lines with only an upper limit is \Pfe (on both gratings that include it), which is formally detected at 1--$2.1\sigma$ but whose spectral appearance is clearly not credible (see \neuI{last figure of Appendix~\ref{sec:alllineprofiles}:} Figure~\ref{fig:lineprofPfe}).
\neuI{All multiply-detected upper limits are very similar between both detectors.}

The only quantitative point of comparison with \citet{luhman23c} concerning the line analysis\footnote{Apart from this, \citet{luhman23c} showed normalised profiles only for \HeIt, \Pag, \Pab, and \Paa.}
is the integrated \Pab line flux. They report $\FLiniemitKont=(6.6\pm1.2)\times10^{-17}~\FEcgs$, which however includes the contribution of the continuum, which we indicate with the $\star$ superscript. %
Our errorbars are similar but only half as large. %
Our continuum-subtracted flux is smaller: $\FPab=(2.7\pm0.6)\times10^{-17}~\FEcgs$.
Taking the continuum to contribute simply $\Delta F=\FKont\times(\Delta \varv)\lambda_0/c$, where $\Delta \varv$ is the fitted width, we obtain 
$\FLiniemitKont=3.8\times10^{-17}~\FEcgs$
or $\FLiniemitKont=5.5\times10^{-17}~\FEcgs$ 
if multiplying $\Delta F$ by $\sqrt{2\pi}$ before adding it to the line excess.  %
Thus we agree in the recovered flux within the errorbars.

We note the following points:
\begin{itemize}
  \item 
The \Paa line is detected with G140H ($R\approx3700$) and G235H ($R\approx2100$), as shown in Figure~\ref{fig:Paa+HeIt}a. The integrated fluxes are almost equal, differing only by the quadratic sum of the $1\sigma$ errorbars. The G235H line is nominally narrower by roughly $2\sigma$ but this reflects the higher fitted continuum level, even though it agrees within less than $1\sigma$ with the continuum fitted on G140H. The true continuum level is unknown, and Figure~\ref{fig:Paa+HeIt}a reveals that some features are absent at lower resolution, but it is unknown whether these features are real or only noise.

  \item
At the other lines covered by two detectors (\Bre, \Brz, \Pfe), the respective continuum level and its uncertainty are the same to much less than 1$\sigma$ between the two detectors. This suggests that the continuum is not heavily variable on a 40-min timescale and that there are no strong systematics between the detectors.

  \item 
For \Pfd, a strong dip near +2200~\kms could lead to an overestimate of the RMS
and could thus reduce the significance of the line.  %

\end{itemize}

\begin{figure}[t]
 \centering
 \includegraphics[width=0.43\textwidth]{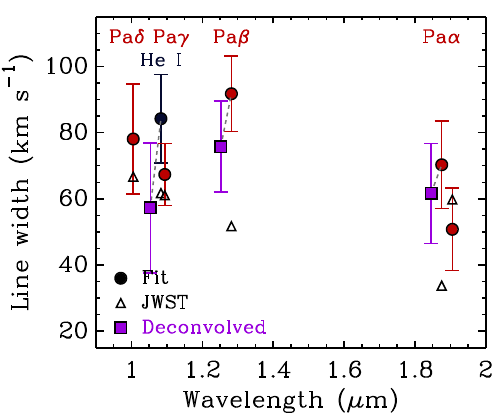}\\
 \includegraphics[width=0.43\textwidth]{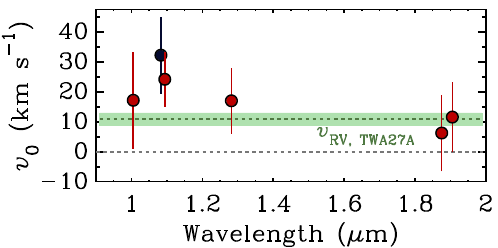}  %
\caption{%
Analysis of the clear detections.
\textit{Top}:
Fitted line widths \Dv (filled circles) and instrumental resolutions \DvInst (open triangles), and deconvolved (intrinsic) line widths \Dvint for \HeIt, \Pab, and $\Paa_1$ (filled squares, connected by dashed lines to \Dv). They are blueshifted and $\Paa_2$ is red-shifted for clarity.
\textit{Bottom}:
Line centroids.
\neuIII{The radial velocity of the primary \citep{faherty16} is shown as a dashed green line with shaded error region.}%
}
\label{fig:Dv+v0}
\end{figure}

In Figure~\ref{fig:Dv+v0}a, we plot the line widths for the clearly-detected lines (\HeIt and the Paschen series).
\Paa from G235H is slightly narrower than the instrumental resolution, but by less than $1\sigma$, which suggests that the continuum level is slightly over- and the line strength underestimated.
However, all other lines (including \Paa from G140H) are wider than the instrumental resolution, which means that they are somewhat resolved.
In fact, assuming that both the intrinsic line shape and the instrumental PSF are Gaussians, we can easily estimate the intrinsic line width (that is, deconvolve) by quadratic subtraction: $\Dvint^2 = (\Delta \varv)^2-\DvInst^2$.
This gives
$\Dvint=(57\pm20)~\kms$ for \HeIt,
$76\pm14$ for \Pab,
$28\pm22$ for \Pag,
$41\pm32$ for \Pad,
$62\pm15$ for $\Paa_1$ (dropping the units),
where the errorbars come from error propagation assuming negligible uncertainty on \DvInst.
Since the fitted line width is less than $1\sigma$ above the instrumental resolution at \Pag and \Pad, we will not consider them for the analysis now.
The inverse-variance weighted average of the deconvolved widths is
$\langle\Dvint\rangle=(67\pm9)~\kms$.
We will discuss this in Section~\ref{sec:disc}.
For reference, the inverse-variance weighted average of the fitted line widths is
$\langle\Delta \varv\rangle=(73\pm5)~\kms$.
The mean deconvolved line width is only $1.5\sigma$ smaller than this.
However, $\langle\Delta \varv\rangle$ is in principle not a meaningful quantity because each line is instrumentally broadened by a different amount. Therefore, we will consider only \Dvint.

The velocity zero-points \neuI{$\varv_0$ (that is, the Doppler offsets from the vacuum central wavelength)} are shown in Figure~\ref{fig:Dv+v0}b.
The \neuI{shorter}-wavelength lines (\HeIt, \Pab, \Pag, \Pad) are less consistent with zero, with an average near $\varv_0\approx+(20\pm10)~\kms$.  %
This, and the $\varv_0$ of the \Paa line from either grating, is close to or consistent with the radial velocity of the primary of $\varv=+(11\pm2)~\kms$ \citep{faherty16}, which is likely the systemic velocity.

\subsection{Accretion luminosity}
 \label{sec:Lacc}

\begin{figure}[t]
 \centering
 \includegraphics[width=0.43\textwidth]{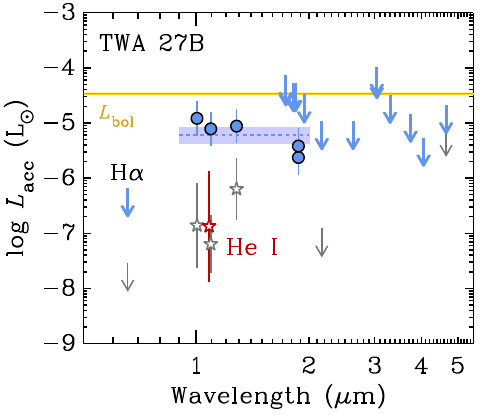}
\caption{%
The accretion luminosity of \twb based on scaling relationships for CTTSs (open grey stars symbols, extrapolated; \citealp{salyk13,alcal17})
and for planets (filled blue circles; \citealp{AMIM21L,ma22}).
Arrows (same colours; thin and thick, respectively) show the $3\sigma$ upper limits of Table~\ref{tab:main},
with \Bra for CTTSs \citep{Komarova+Fischer2020} outside the plot
because it predicts $\LAkk\sim10^{-10}~\LSonne$.
The \LAkk from \HeIt (dark red for clarity) is from \citet{alcal14}, without a planetary-scaling counterpart.
We display the weighted mean of the \LAkk inferred from planetary-shock-model scalings (blue dashed line and band). The bolometric luminosity (gold; from \citealt{luhman23c}) is shown for comparison.
}
\label{fig:LAkks}
\end{figure}

In Figure~\ref{fig:LAkks}, we compare the accretion luminosity \LAkk derived from each line independently.
We use \neuI{the \Ha upper limit (see Appendix~\ref{sec:HSTHa}),} 
$F<1.6\times10^{-17}~\FEcgs$, which we multiply by three to estimate the $3\sigma$ limit.
We use $\LAkk(\LLinie)$ relationships
\neuI{of the form $\log_{10}\LAkk/\LSonne=a\log_{10}\LLinie/\LSonne+b$, where $(a,\,b)$ are fit coefficients.}
As in \citet{betti22b,betti22c}, we use both
the relationships for CTTSs 
extrapolated down to the line luminosities measured at \twb, %
and also the $\LAkk(\LLinie)$ relationships based on detailed shock models designed for planetary-mass objects \citep{AMIM21L,ma22}. 
We will refer to \neuI{the latter} as ``planetary \LAkk''.
Magnetospheric accretion columns could contribute to the line fluxes also for planetary-mass objects but these models do not include this.
We note that the \citet{aoyama18} model does not make predictions for helium lines.  %

\neuI{Specifically, for the CTTS scalings we use the coefficients in \citet{alcal17} for all lines available,
complementing with
\citet{salyk13} for \Pab, \citet{alcal14} for \HeIt, and \citet{Komarova+Fischer2020} for \Bra. Most coefficients are summarised in \citet{AMIM21L} and compared to the planetary values.}
We note that no CTTS scalings exist for \Paa, \Brb, \Brd, and \Pfd.

We propagate the error from the fit and the uncertainty on the line flux.
For CTTSs, the fit error is from the errors on $a$ and $b$,
and for the planetary relationships we fix $\sigma=0.3$~dex (see \citealp{AMIM21L}),
added through error propagation to $\sigma$ from the line flux.

The inverse-variance-weighted average of the planetary \LAkk from 
the detections at \Paa--\Pad is $\langle\log(\LAkk/\LSonne)\rangle=-5.23\pm0.14$~dex (blue dashed line and 1$\sigma$ band in Figure~\ref{fig:LAkks}).
Even though the errorbars are small ($\sigma\approx0.2$~dex), the \LAkk values from the individual lines lie within $1\sigma$ of the average.
With the exception of \Ha, to which we will return below, the upper limits from the other lines are consistent with this.
When using the extrapolated CTTS scalings, the errorbars on \LAkk are larger ($\sigma\approx0.6$~dex), but there too, the inferred \LAkk from each line agrees within $1\sigma$ with the average $\langle\log(\LAkk/\LSonne)\rangle\approx-7$~dex (not shown). The major difference is that this is roughly 1.7~dex smaller than the average \LAkk as inferred from planetary models. This is in line with the conclusions of \citet{AMIM21L} or, for \Dlrmb, \citet{betti22b,betti22c}.

Interestingly, the \LAkk corresponding to the $3\sigma$ upper limit on \FHa is lower than the average \LAkk by 1~or 0.5~dex according to the planetary or CTTS relationships, respectively. This apparent discrepancy could be explained by time variability (the \Ha was measured in 2011, the other lines in 2023),
wavelength-dependent absorption due to dust (see Figure~9b of \citealt{maea21}, but given the low accretion rate, \neuI{absorption} is likely \neuI{negligible}), or differences in the emission mechanism for \Ha compared to the other lines (at least for TW~Hydra, \Ha does not correlate as well as the other lines with \Lacc; \citealp{herczeg23}). No strong arguments can be made for or against these factors. Otherwise, instrumental or other observational effects (including systematics) need to be invoked, but we assume that they cannot explain most of the 0.5--1~dex difference.

In Figure~\ref{fig:LAkks}, we also applied the extrapolated \HeIt $\LAkk(\LLinie)$ scaling relationship for CTTSs, which was studied only by \citet{alcal14}.
The helium triplet is very sensitive to both winds and accretion \citep[e.g.,][]{fischer08,thanathibodee22,erkal22} and should be used with caution, as \citet{alcal14} note.  %
The modest resolution of the NIRSpec data, $R=2062$, does not allow for detailed studies of the line shape (see the types proposed in \citealt{thanathibodee22}), especially since the line width is barely wider than the instrumental broadening (Figure~\ref{fig:Dv+v0}a; profile in Fig.~\ref{fig:lineprofHeIt}). Nevertheless, the \LAkk inferred from \HeIt is consistent with the average from the other lines. High-resolution line profiles could reveal whether this is a coincidence, and theoretical predictions extending the \citet{aoyama18} models would be welcome.

\subsection{Accretion rate}

Combined with an inferred mass of $\MP\approx5~\MJ$ and $\RP=1.4~\RJ$ from \citet{luhman23c},
the average \LAkk can be translated into an ``accretion rate'' \MPkt, whose meaning requires discussion (see afterwards).
From the planetary scalings, we obtain
$\MPkt\approx\LAkk/(G\MP/\RP) \approx 5\times10^{-9}~\MPktEJ$.
Using instead the extrapolation of the \citet{alcal17} CTTS relationships would imply $\MPkt\approx0.1\times10^{-9}~\MPktEJ$,
while the scaling of \citet{natta04} applied to our \FPab value yields
$\MPkt\approx0.3\times10^{-10}~\MPktEJ$.
This is slightly lower than the $\MPkt\sim10^{-10}$--$10^{-9}~\MPktEJ$ that \citet{luhman23c} report using that relationship.
The reason is that these scalings are defined for continuum-subtracted (i.e., photospheric-emission-corrected) values, while \citet{luhman23c} took the total line flux, leading to an overestimate by
around 0.4~dex. This is comparable to the scatter in these relationships
and clearly smaller than their systematic uncertainties \citep{betti23},
and thus not a major issue.
The targets analysed in \citet{alcal17} have \Pab luminosities $\LLinie\sim10^{-6.5}$--$10^{-3}~\LSonne$.
Thus, applying this relationship to \twb requires an extrapolation by 2~dex. This is only slightly smaller than the range of data and therefore possibly acceptable, but it would require some validation.

This accretion rate, whether using the planetary scalings or the extrapolated CTTS ones, is only an estimate due to uncertainties in the prefactors (for instance due to a finite starting radius for the infall).
However, even the presumably more realistic, and much higher, value from the planetary scalings implies an ``accretion rate'' that is small in the sense that the nominal mass doubling time
$\tau=\MP/\MPkt\sim1$~Gyr, which is orders of magnitude longer than the age of the system ($10\pm2$~Myr; \citealt{luhman23b}) or of the TWA.
Thus \twb would be at the very end of its formation.
This is in line with the finding that \LAkk is smaller than the bolometric luminosity (see Figure~\ref{fig:LAkks}) by about 0.4~dex according to planetary scalings, and even more if extrapolating the CTTS scalings. While of interest for observers and theoreticians alike, the accretion processes seem to be currently of very subdominant existential relevance for \twb.

However, as discussed in \citet{m22Schock}, this \MPkt is not necessarily the growth rate of the planet but rather only what is hitting the planetary surface and/or the CPD close to the planet: in practice, gas needs to shock at $\varv\gtrsim30~\kms$ to emit lines\footnote{This applies to molecular hydrogen, since it first needs to dissociate, using up energy that could otherwise go into line emission. Atomic hydrogen could emit for lower shock velocities, but at low shock velocities the accreting hydrogen is expected to be coming in molecular form (see Figure~12b of \citealt{Aoyama+2020}).} \citep{aoyama18}. \twb might also be accreting through a boundary layer, which \neuI{likely} does not generate lines. Then, the total accretion rate would be larger than inferred. Given the mass ratio with \twa ($q\approx0.2$) and their separation (55~au), it is unclear whether physical scales comparable to or larger than the Hill sphere $\RH\approx22$~au are feeding \twb, or whether it can draw the matter it is accreting only from a CPD. There is no clear \neuI{published} IR flux excess \neuI{(while, as \citealt{luhman23c} note, there might be in unpublished 5--28~\mum MIRI data)},
and the Atacama Large Millimeter/submillimeter Array (ALMA) upper limit from \citet{ricci17} constrains little the amount of mass available in the system or the size of the mass reservoir.
On the other hand, if \twb is undergoing magnetospheric accretion and the columns contribute to the line emission \citep{hartmann16}, the line flux may indeed be tracing \neuI{(nearly)} the whole mass flow.

\begin{figure}[t]
 \centering
 \includegraphics[width=0.43\textwidth]{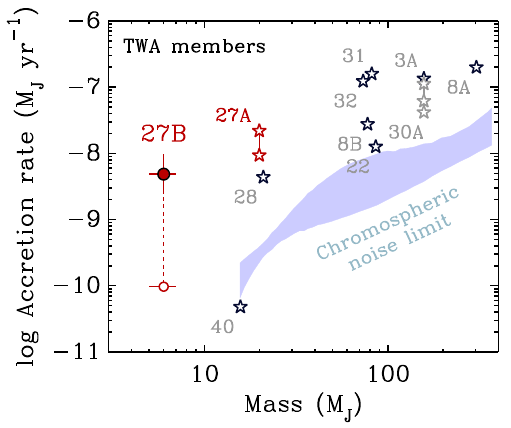}
\caption{%
Accretion rates of TWA objects.
We show the data of \citet{venuti19} (grey stars; multiple observations are joined by a line; only TWA~1 is off the plot with $M\approx600~\MJ$)
and add our analysis of \twb
using the \citet{AMIM21L} relationships (filled circle)
or the extrapolated CTTS relationships (open circle).
Labels give the TWA numbers.
The blue band indicates the chromospheric noise limit at 3--10~Myr from \citet{venuti19}.
}
\label{fig:MPkt}
\end{figure}

We keep these caveats in mind but compare in Figure~\ref{fig:MPkt} this accretion rate with those of the other TWA members.
The \MPkt values are from \citet{venuti19}, with TWA~30B excluded because it is likely severely affected by extinction \citep{looper10b,rodriguez15}.
Especially towards higher-mass objects, the line luminosities might trace the entire mass flow,
whereas for the lower mass objects (\twa and TWA~28, since no lines were detected at TWA~40), the reported \MPkt values might be lower limits.

Intriguingly, the accretion rate onto \twb according to the planetary scaling relationship is somewhat smaller than the CTTS-scalings-based \MPkt for \twa, with an instantaneous ratio $\eta_{\mathrm{inst}}\equiv\MPktB/(\MPktA+\MPktB)
\approx0.2$,
where $\MPktA\approx10^{-7.8}~\MPktEJ$ \citep{venuti19}.
At a mass ratio $q\approx 0.2$,
different hydrodynamics simulations predict on average a steady-state $\eta\approx0.6$--0.9 \citep{bate00,lai23}.  %
That the theoretical and observational $\eta$ are only within a factor of a few from each other suggests that the \MPkt estimate using the planetary paradigm for \twb might be relatively accurate.
However, the measurements are not contemporaneous (\twa: taken in 2010 and 2012; \twb: 2023), while accretion-rate variability is expected in binary systems on several timescales (e.g., \citealp{mu20}). This comparison should therefore be repeated after further monitoring.

\section{Discussion}
 \label{sec:disc}

We briefly discuss some aspects of our results.

\textit{Line width.}---%
We found an average deconvolved (intrinsic) line width $\Dvint=(67\pm9)~\kms$ based on the well-resolved lines \HeIt, \Paa, and \Pab. The other clearly-detected lines, \Pag and \Pad, have a fitted width barely wider than the instrumental resolution.
Shock models predict the intrinsic line width to be not directly equal to but of the order of the preshock velocity \citep{aoyama18,Aoyama+Ikoma2019}.
The exact line width depends on the preshock conditions because they set where (at what depth and thus temperature) in the postshock region a particular line is formed.
Hydrogen and helium lines might originate from different depths given the different excitation energies of the starting levels.
Nevertheless, all intrinsic line widths and the average are consistent with each other to within $1\sigma$.

If $\MP=5~\MJ$ and $\RP=1.4~\RJ$ \citep{luhman23c}, the surface freefall velocity is $\vFfinfty=117~\kms$, so that $\Dvint\approx0.6\vFfinfty$.
Qualitatively,  %
this therefore seems to be in line with shock-model predictions.
This also agrees with simulations of accretion onto gap-opening planets, which show that while most gas falling from Hill-sphere scales lands on the CPD and not on the planetary surface \citep{tanigawa12}, the contribution from the planetary surface should dominate \citep{m22Schock}.
At the same time, this line width also seems broadly consistent with predictions from magnetospheric accretion \citep{thanathibodee19}.
Thus, only a \neuI{medium}-resolution line profile is not sufficient to distinguish the two scenarios, and both higher-resolution observations and quantitative modelling are required \citep{demars23}.
However, this \Dvint matches well the interpretation that \twb is not a higher-mass (substellar) object with an inclined disc leading to high extinction (see \citealp{luhman07d}) since the higher mass would likely lead to a wider line than observed.

\textit{Chromospheric activity?}---%
Accretion at low-mass objects is a barely-charted territory and a valid concern is whether the observed line emission comes from accretion or chromospheric activity. Figure~\ref{fig:MPkt} shows that the \MPkt of \twb is 1.5~dex above the chromospheric noise limit \citep{manara13,manara17b,venuti19} at $\MP=20~\MJ$. How the limit behaves at lower masses, that is, closer to \twb's $\MP\approx5~\MJ$, is an open question; it might extrapolate as a powerlaw or drop precipitously below a certain mass. However, a sharp rise by several orders of magnitude seems unlikely. Therefore, despite the statistical uncertainties in \MPkt, accretion appears to be the likeliest source of the lines at \twb. A similar argument can be made for \Dlrmb \citep{eriksson20,betti22b,betti22c,ringqvist23}.

\textit{Helium line emission}---%
With its clear \HeItVak emission \citep{luhman23c}, \twb joins the select club of known accretors below $\approx20~\MJ$ exhibiting \HeI lines.
To the best of our knowledge,
the other members are \Dlrmb
(\neuI{$\MP\approx12~\MJ$;} detections at
$\lambda\lambda6678$, 7065, 7281, 10833; \citealp{eriksson20,betti22b})
and 2MASS J11151597+1937266\footnote{\neuI{While \citet{theissen18} qualified the hydrogen-line emission as coming from persistent magnetic activity or weak accretion, they were assuming a distance $d=(37\pm6)$~pc. The updated $d=(45\pm2)$~pc \citep{gEDR3} %
might imply that accretion is stronger than initially thought, and the mass somewhat higher, but this needs a quantitative re-assessment.}}
(\neuI{$\MP\approx7$--$21~\MJ$}; detections at $\lambda\lambda4471$, 5876, 6678, 7065;
\citealp{theissen17,theissen18}),
with the next least massive objects around
$\MP\approx20~\MJ$ \citep{mohanty05,herczeg09}.
As noted \neuI{in Section~\ref{sec:Lacc}}, interpreting the \HeIt line shape at stellar accretors is difficult \citep{kwan07,erkal22}, and the situation for planetary-mass accretors is unknown.
Higher-resolution observations should allow disentangling possible contributions from a CPD wind, accretion funnels, or post-shock emission.

\section{Summary and conclusion}
 \label{sec:summconc}

We have re-analysed the processed JWST/NIRSpec spectrum of \twb presented in \citet{luhman23c}, focussing on the accretion indicators because they stated detections of three or four Paschen lines and of the \HeIt triplet. We searched systematically for all accessible hydrogen lines, fitted and subtracted the continuum, and measured line shapes and total fluxes.
We quantified the uncertainty on the line shape and integrated flux by measuring the ``photospheric noise'', which is the continuum residual from the (not-modelled) atmosphere.
Our results are:
\begin{enumerate}
  
  \item \HeIt is detected at $>4\sigma$ and the Paschen-series lines that NIRSpec covers (\Paa--\Pad) at (3.5--$5)\sigma$. These are robust detections. The \Bra, \Brb, and \Brg lines are tentatively detected with (2--3)$\sigma$. The \Pfg and \Pfd signals are marginal. For the other lines (\Brd--\Brz, \Pfb, and \Pfe), we obtain upper limits.

  \item The \Paa line is covered by two grating-and-filter combinations, which observed \twb $\sim40$~min apart. The similarity of the continuum and the line shape between both observations suggests little continuum variability and no strong systematics between the gratings.

  \item 
  We independently fit each line and find that for the
  lines clearly detected at $>3\sigma$, namely \HeIt, $\Paa_1$, and \Pab, the fitted widths are well above the instrumental resolution (especially the latter two thanks to $R\approx3800$ and $\approx2500$, respectively).
  Their de-convolved (intrinsic) widths are consistent with each other and averaging to $\Dvint=(67\pm9)~\kms$. This is around 60\,\% of the free-fall velocity at the surface of \twb.
  The width is qualitatively consistent with shock-model predictions \citep{aoyama18} and the result that the planetary-surface shock and not the CPD-surface shock should dominate the emission \citep{m22Schock}, but also with expectations from magnetospheric accretion \citep{thanathibodee19}.
  Detailed modelling is required to relate preshock velocities and line widths and to help determine where and how the emission lines are formed.
 
  \item We find a \Pab line-excess flux $\FPab=(2.7\pm0.6)\times10^{-17}~\FEcgs$.
  Adding the contribution from the continuum,
  we recover to $1\sigma$ the total line flux reported by \citet{luhman23c},
  $(6.6\pm1.2)\times10^{-17}~\FEcgs$, \neuI{which includes the continuum}.

  \item Using scalings derived for CTTS (e.g., \citealp{alcal17}) extrapolated to planetary luminosities or scalings based on detailed models yields accretion luminosities that are consistent between the different lines but discrepant by 1.7~dex between the two approaches, as found before \citep{betti22b,betti22c}.
  Correspondingly, the line-producing (shock) gas mass flow rate $\MPkt\approx5\times10^{-9}~\MPktEJ$ is about 50~times higher according to planetary scalings than with the extrapolated CTTS relationships. If \MPkt is the growth rate of \twb, its formation is over. The possibility of magnetospheric accretion and other considerations introduce uncertainties about the meaning of \MPkt.
  
  \item
  All integrated line luminosities are on the order of $\LLinie\sim10^{-9}~\LSonne$. Despite theoretical uncertainties, the lines detected come robustly from accretion processes and not from chromospheric activity.
  
\end{enumerate}

A detailed study of the atmospheric properties could help identify atomic and molecular features in the spectrum of \twb. Removing them would reduce the continuum noise, improve the accuracy of the inferred line shape and flux, and increase the statistical significance by decreasing the noise. Thus the significance of our detections might be currently underestimated. Deeper and spatially better resolved ALMA observations would be very valuable to constrain the amount of mass available in the TWA~27 system and the location of the gas reservoir from which \twb is drawing.

Twenty years after its discovery, \twb still holds many surprises.
It could well be that also other young directly-imaged companions are accreting even though no CPD has been detected so far and even if the parent disc is long gone. A deep look with JWST could be worthwhile,
and our results suggest that \HeIt, \Pab, and \Paa might be particularly well-suited tracers.

\begin{acknowledgments}
{\small
We are indebted to Kevin Luhman for impressively fast and helpful comments and answers and for generously sharing his data. We thank Tomas Stolker for help with the flux errorbars; Paul Molli\`ere and Gabriele Cugno for comments on JWST; and Sarah Betti for discussions of upper limits.

G-DM acknowledges the support of the DFG priority program SPP 1992 ``Exploring the Diversity of Extrasolar Planets'' (MA~9185/1).
G-DM also acknowledges the support from the Swiss National Science Foundation under grant
200021\_204847
``PlanetsInTime''.
JH is supported by JSPS KAKENHI Grant Numbers 21H00059, 22H01274, and 23K03463.
YA is funded by China Postdoctoral Science Foundation (2023M740110).
Parts of this work have been carried out within the framework of the NCCR PlanetS supported by the Swiss National Science Foundation.
\neuI{%
This work is based in part on observations made with the NASA/ESA/CSA James Webb Space Telescope. The data were obtained from the Mikulski Archive for Space Telescopes at the Space Telescope Science Institute, which is operated by the Association of Universities for Research in Astronomy, Inc., under NASA contract NAS 5--03127 for JWST. These observations are associated with program GTO~1270.
This research is based in part on observations made with the NASA/ESA Hubble Space Telescope obtained from the Space Telescope Science Institute, which is operated by the Association of Universities for Research in Astronomy, Inc., under NASA contract NAS 5--26555. These observations are associated with program~12225.
Some of the data presented in this paper can be obtained from the Mikulski Archive for Space Telescopes (MAST) at the Space Telescope Science Institute. The specific HST and JWST observations analysed can be accessed via \dataset[DOI: 10.17909/2f9b-ea80]{http://dx.doi.org/10.17909/2f9b-ea80}.%
}%
This research has made use of NASA's Astrophysics Data System Bibliographic Services.

Figures~\ref{fig:Paa+HeIt}--\ref{fig:MPkt} were produced using \href{https://github.com/gnudatalanguage/gdl}{\texttt{GDL}}, an actively-developed open-source drop-in alternative to \texttt{IDL}.
\neuI{Figure~\ref{fig:Ha2D} was made with \url{https://github.com/AstroJacobLi/smplotlib}, a package that (finally!) makes easily available in \texttt{python} the Hershey fonts of good ol' \texttt{SuperMongo} and \texttt{IDL}.} %
Figures in Appendix~\ref{sec:alllineprofiles} used \texttt{gnuplot} with the terminal \texttt{pdfcairo} and the font Priori~Sans. % (discovered thanks to \url{https://www.tjodsavnid.fo}).

}
\end{acknowledgments}

\appendix

\section[HST data at Halpha]{HST data at \Ha}
 \label{sec:HSTHa}

\begin{figure}[t]
 \centering
 \includegraphics[width=0.43\textwidth]{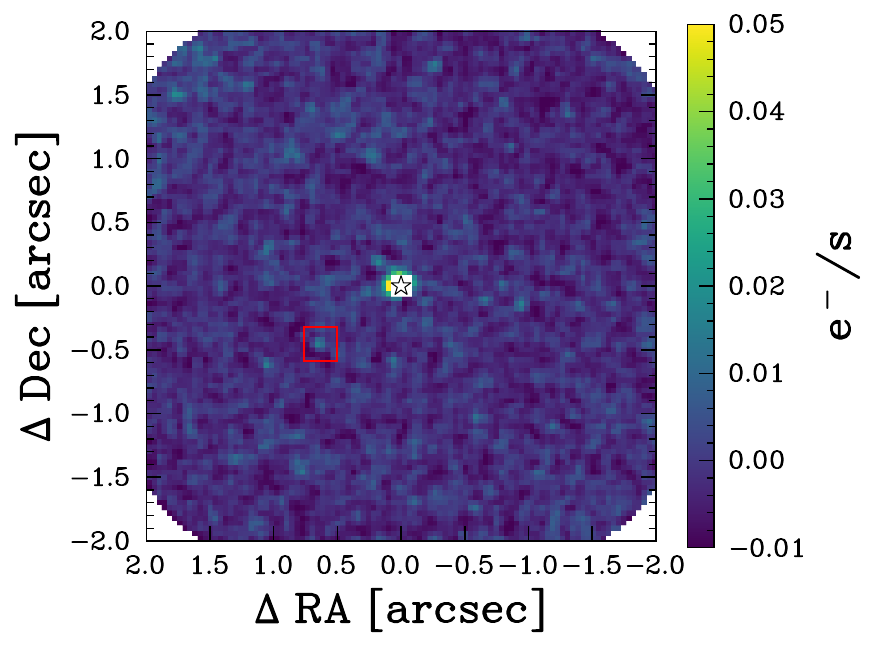}
\caption{%
\neuI{HST \Ha (WFC3/UVIS2/F656N; 
$\Delta\lambda=18$~\AA, $R=372$, $\Dv=c/R=807~\kms$)
view of the TWA~27 system, with subtracted primary (star symbol) and non-detected companion (red box).%
}%
}
\label{fig:Ha2D}
\end{figure}

In Figure~\ref{fig:Ha2D}, we show the primary-subtracted F656N image \neuII{(that is, at \Ha)} of the TWA~27 system taken by the Hubble Space Telescope with WFC3/UVIS2.
TWA~27 was observed on 2011 March 28 (PID: 12225, PI: A.~Reiners), with an integration time of 120~s.
The location of \twb is indicated by a red box.
The flux at the position of \twb, measured in an $r=2.5$~pixel aperture, is $2.2\sigma$ above the background,
which was calculated as the standard deviation from $r=2.5$-pixel apertures at the same angular separation as \twb.
\neuII{Therefore, the small excess is tantalising but} a confident \Ha detection ($>3\sigma$) would require additional observations.
We use the flux upper limit in Section~\ref{sec:Lacc}.

\section{All line profiles}
 \label{sec:alllineprofiles}

\begin{figure*}[t]
 \centering
 \includegraphics[width=0.97\textwidth]{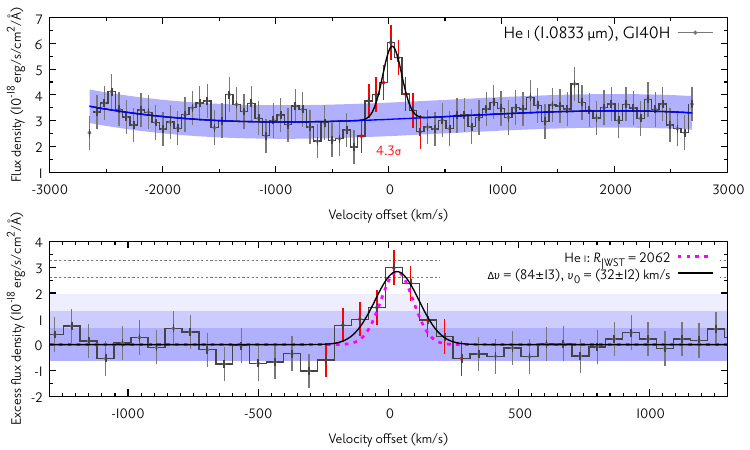}
\caption{%
Line profile for \HeIt. We show the fitted continuum and line (black) and just the continuum (blue line). \textit{Top}: whole range used for calculating the continuum, excluding the points in red (close to the line centre). Blue band: $\pm1\sigma$ range (RMS of continuum). \textit{Bottom}: zoom-in on the continuum-subtracted line. Bands: (1, 2, $3)\sigma$, dotted grey lines: (4, $5)\sigma$.
Pink: instrumental broadening. The dark grey part of the errorbars: only the error on the bin (as we re-determined it); full errorbar: adding the continuum RMS in quadrature.
}
\label{fig:lineprofHeIt}
\end{figure*}

\begin{figure*}
 \centering
 \includegraphics[width=0.97\textwidth]{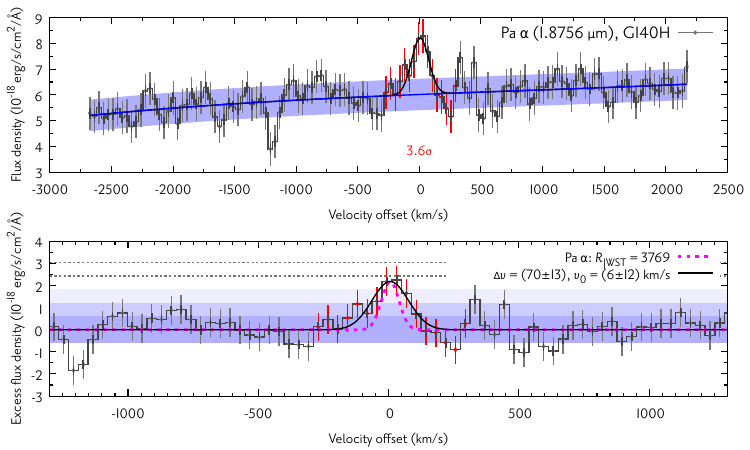}
 \includegraphics[width=0.97\textwidth]{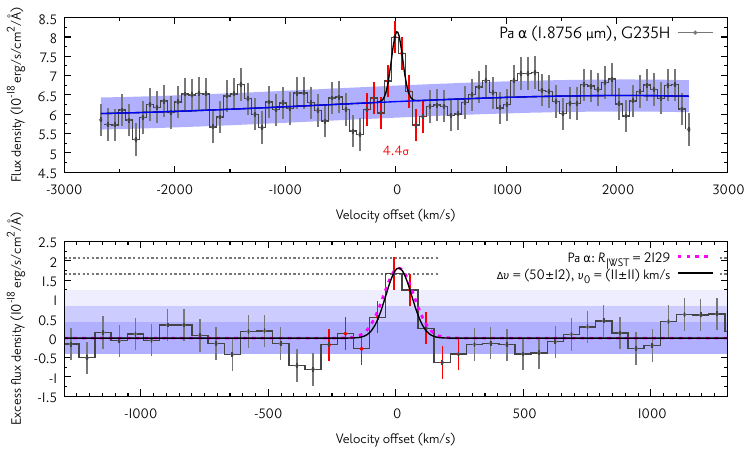}
\caption{%
As in Figure~\ref{fig:lineprofHeIt} but for \Paa on the two gratings: red half of G140H (top), blue half of G235H (bottom).
}
\label{fig:lineprofPaa2}
\end{figure*}

\begin{figure*}
 \centering
 \includegraphics[width=0.97\textwidth]{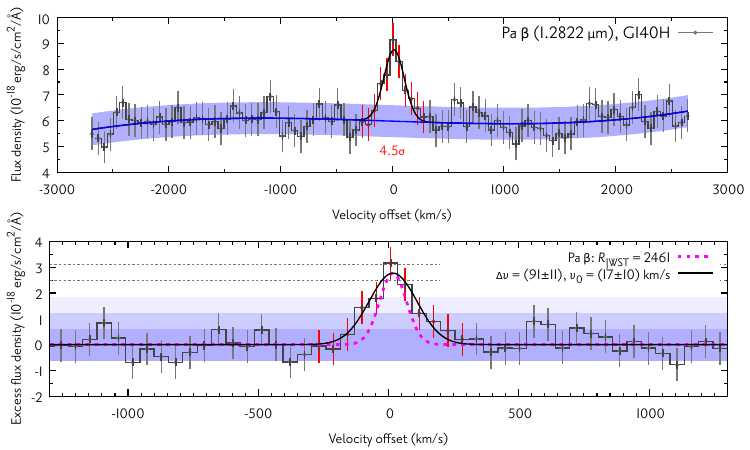}
 \includegraphics[width=0.97\textwidth]{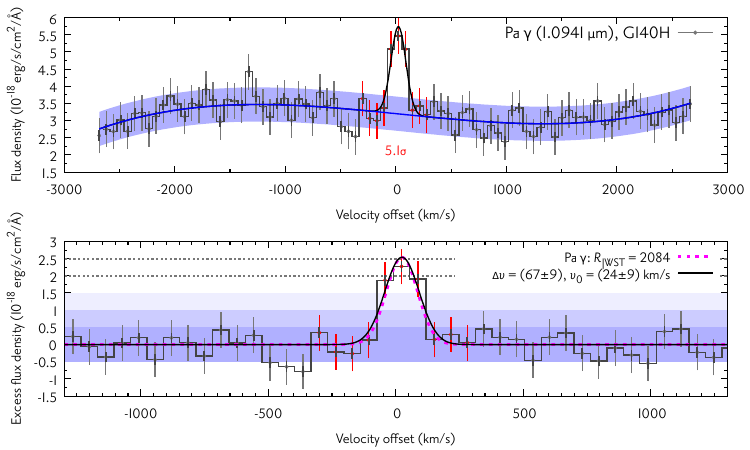}
\caption{%
As in Figure~\ref{fig:lineprofHeIt} but for \Pab and \Pag.
}
\label{fig:lineprofPabPag}
\end{figure*}

\begin{figure*}
 \centering
 \includegraphics[width=0.97\textwidth]{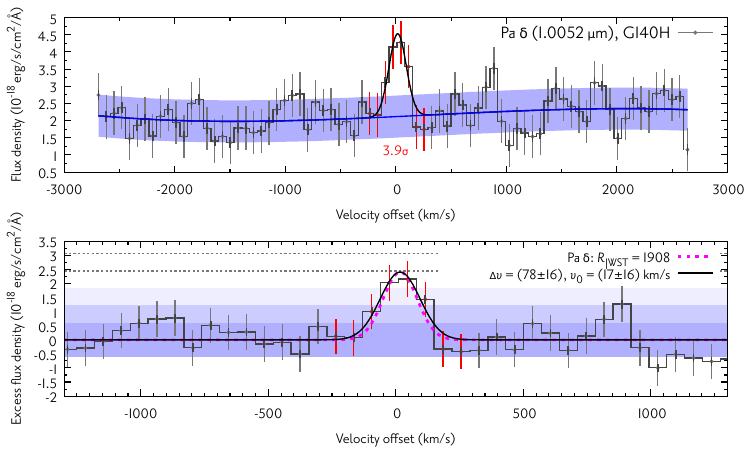}
\caption{%
As in Figure~\ref{fig:lineprofHeIt} but for \Pad, the highest-order potentially detectable Paschen line. 
}
\label{fig:lineprofPad}
\end{figure*}

\begin{figure*}
 \centering
 \includegraphics[width=0.97\textwidth]{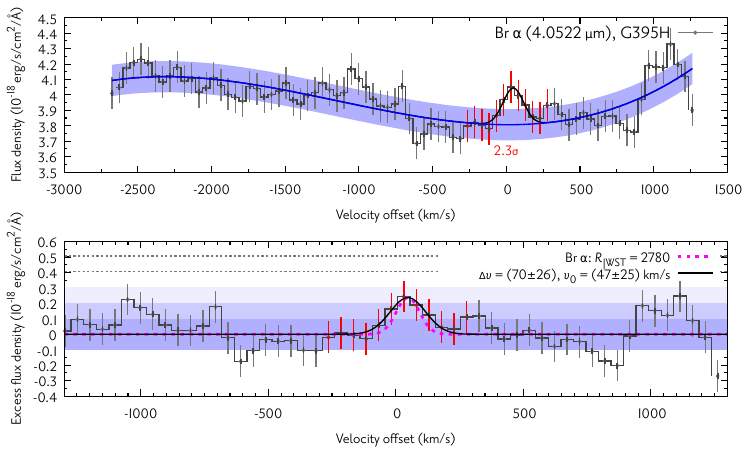}
 \includegraphics[width=0.97\textwidth]{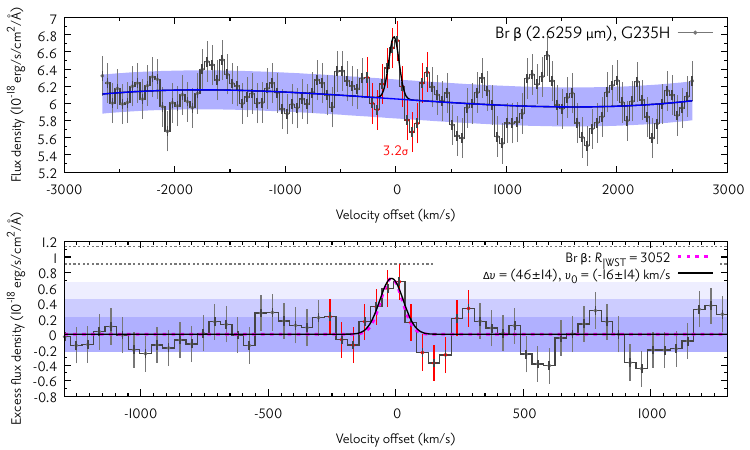}
\caption{%
As in Figure~\ref{fig:lineprofHeIt} but for \Bra and \Brb.
}
\label{fig:lineprofBraBrb}
\end{figure*}

\begin{figure*}
 \centering
 \includegraphics[width=0.97\textwidth]{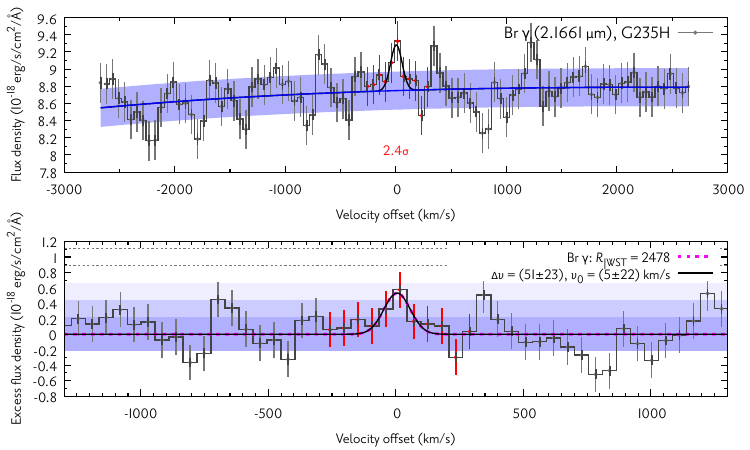}
 \includegraphics[width=0.97\textwidth]{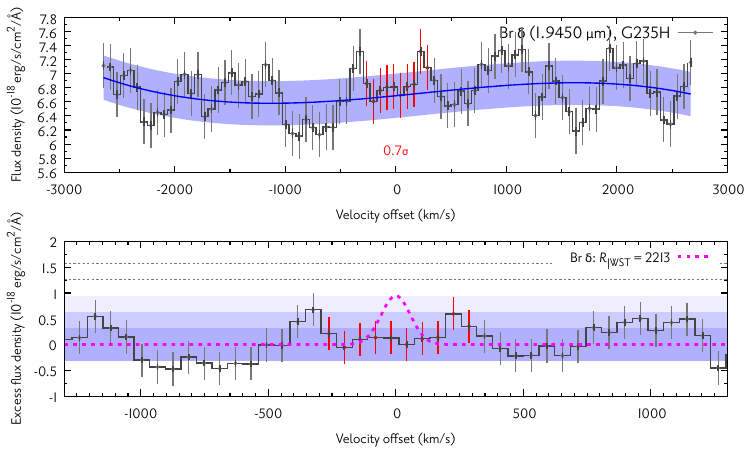}
\caption{%
As in Figure~\ref{fig:lineprofBraBrb} but for \Brg, which is tentatively detected, and for \Brd, which is a clear non-detection. For non-detections, the height of the instrumental Gaussian is set to $3\sigma$ for illustration purposes.
}
\label{fig:lineprofBrgBrd}
\end{figure*}

\begin{figure*}
 \centering
 \includegraphics[width=0.97\textwidth]{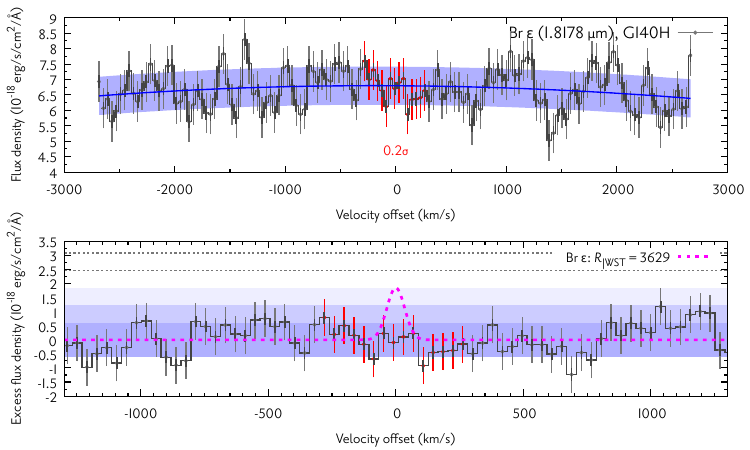}
 \includegraphics[width=0.97\textwidth]{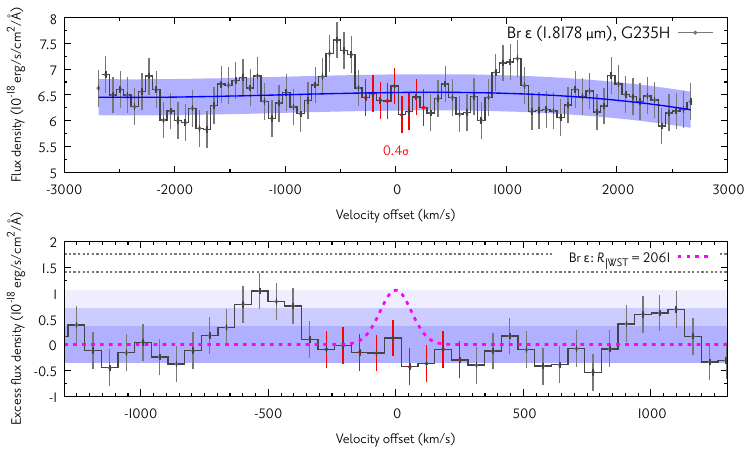}
\caption{%
As in Figure~\ref{fig:lineprofBraBrb} but for \Bre (clearly non-detected) through two gratings (top: G140H, bottom: G235H).
}
\label{fig:lineprofBre2}
\end{figure*}

\begin{figure*}
 \centering
 \includegraphics[width=0.97\textwidth]{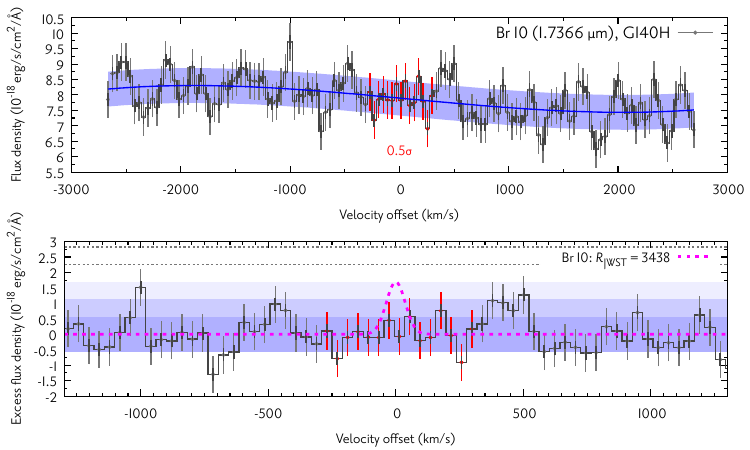}
 \includegraphics[width=0.97\textwidth]{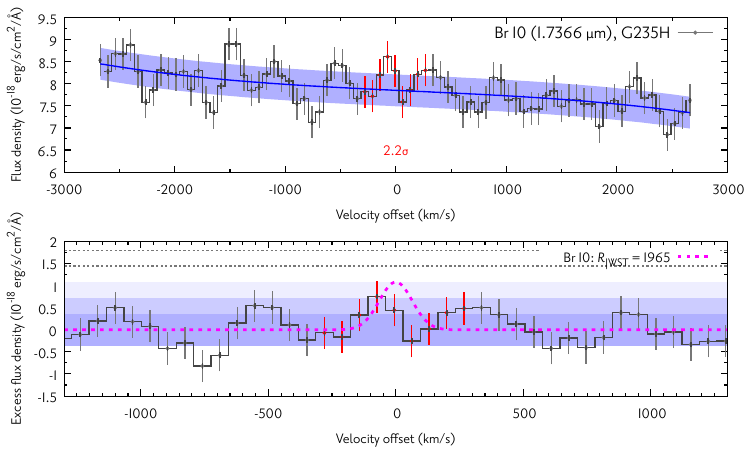}
\caption{%
As in Figure~\ref{fig:lineprofBre2} but for \Brz (clearly non-detected) through two gratings (top: G140H, bottom: G235H).
}
\label{fig:lineprofBrz2}
\end{figure*}

\begin{figure*}
 \centering
 \includegraphics[width=0.97\textwidth]{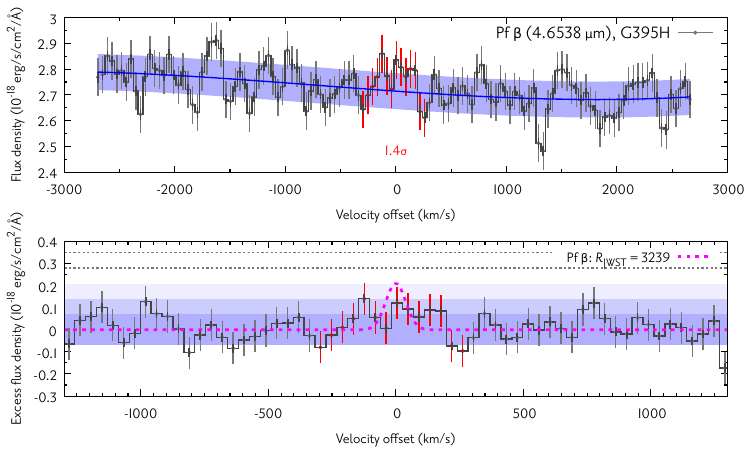}
 \includegraphics[width=0.97\textwidth]{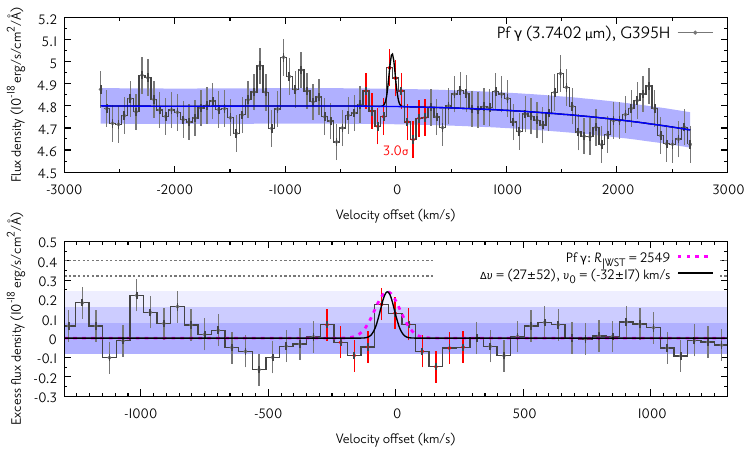}
\caption{%
As in Figure~\ref{fig:lineprofBraBrb} but for \Pfb (not detected) and \Pfg (tentatively detected, but with more significance if the continuum is in fact lower).
}
\label{fig:lineprofPfbPfg}
\end{figure*}

\begin{figure*}
 \centering
 \includegraphics[width=0.97\textwidth]{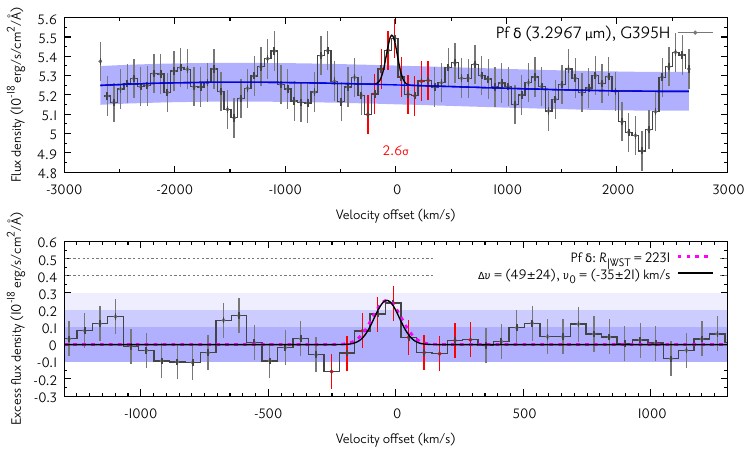}
\caption{%
As in Figure~\ref{fig:lineprofPfbPfg} but for \Pfd, very tentatively detected.
}
\label{fig:lineprofPfd}
\end{figure*}

\begin{figure*}
 \centering
 \includegraphics[width=0.97\textwidth]{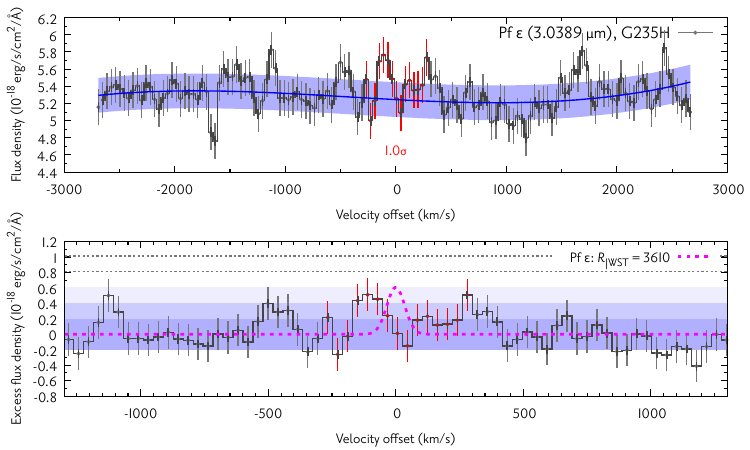}
 \includegraphics[width=0.97\textwidth]{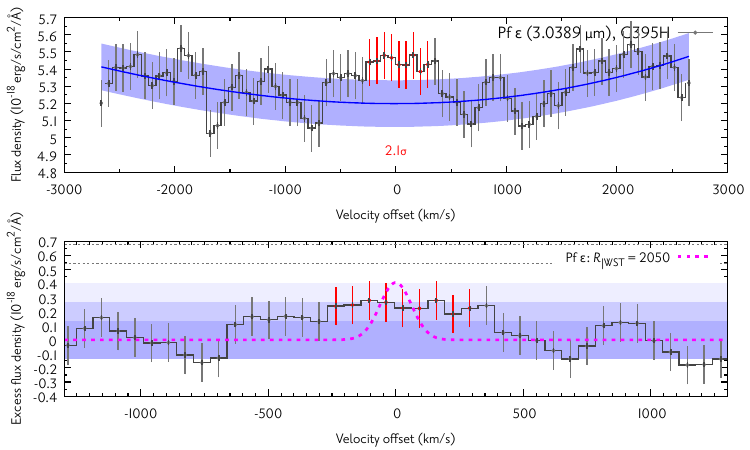}
\caption{%
As in Figure~\ref{fig:lineprofPfbPfg} but for \Pfe from two gratings (red half of G235H on top, blue half of G395H on the bottom), in both not detected.
}
\label{fig:lineprofPfe}
\end{figure*}

In Figures~\ref{fig:lineprofHeIt}--\ref{fig:lineprofPfe} we show profiles of the \HeIt and the hydrogen lines for which we \neuI{searched}. In each case we show a broader region with the fitted continuum (top panel) and the continuum-subtracted profile
in a zoomed-in region (bottom).
\neuI{The integrated significance of each line, using $N_\sigma^F$ from Table~\ref{tab:main}, is indicated in red below each top-panel curve.}

~  %

\bibliography{std}{}

\begin{thebibliography}{}
\providecommand\natexlab[1]{#1}
\providecommand\JournalTitle[1]{#1}

\bibitem[{{Alcal{\'a}} {et~al.}(2014){Alcal{\'a}}, {Natta}, {Manara}, {Spezzi},
  {Stelzer}, {Frasca}, {Biazzo}, {Covino}, {Randich}, {Rigliaco}, {Testi},
  {Comer{\'o}n}, {Cupani}, \& {D'Elia}}]{alcal14}
{Alcal{\'a}}, J.~M., {Natta}, A., {Manara}, C.~F., {et~al.} 2014,
  \href{http://dx.doi.org/10.1051/0004-6361/201322254}{\JournalTitle{\aap},
  561, A2}

\bibitem[{{Alcal{\'a}} {et~al.}(2017){Alcal{\'a}}, {Manara}, {Natta}, {Frasca},
  {Testi}, {Nisini}, {Stelzer}, {Williams}, {Antoniucci}, {Biazzo}, {Covino},
  {Esposito}, {Getman}, \& {Rigliaco}}]{alcal17}
{Alcal{\'a}}, J.~M., {Manara}, C.~F., {Natta}, A., {et~al.} 2017,
  \href{http://dx.doi.org/10.1051/0004-6361/201629929}{\JournalTitle{\aap},
  600, A20}

\bibitem[{{Aoyama} \& {Ikoma}(2019)}]{Aoyama+Ikoma2019}
{Aoyama}, Y., \& {Ikoma}, M. 2019,
  \href{http://dx.doi.org/10.3847/2041-8213/ab5062}{\JournalTitle{\apjl}, 885,
  L29}

\bibitem[{{Aoyama} {et~al.}(2018){Aoyama}, {Ikoma}, \& {Tanigawa}}]{aoyama18}
{Aoyama}, Y., {Ikoma}, M., \& {Tanigawa}, T. 2018,
  \href{http://dx.doi.org/10.3847/1538-4357/aadc11}{\JournalTitle{\apj}, 866,
  84}

\bibitem[{{Aoyama} {et~al.}(2021){Aoyama}, {Marleau}, {Ikoma}, \&
  {Mordasini}}]{AMIM21L}
{Aoyama}, Y., {Marleau}, G.-D., {Ikoma}, M., \& {Mordasini}, C. 2021,
  \href{http://dx.doi.org/10.3847/2041-8213/ac19bd}{\JournalTitle{\apjl}, 917,
  L30}

\bibitem[{{Aoyama} {et~al.}(2020){Aoyama}, {Marleau}, {Mordasini}, \&
  {Ikoma}}]{Aoyama+2020}
{Aoyama}, Y., {Marleau}, G.-D., {Mordasini}, C., \& {Ikoma}, M. 2020,
  \href{http://dx.doi.org/10.48550/arxiv.2011.06608}{\JournalTitle{arXiv
  e-prints}, arXiv:2011.06608}

\bibitem[{{Bate}(2000)}]{bate00}
{Bate}, M.~R. 2000,
  \href{http://dx.doi.org/10.1046/j.1365-8711.2000.03333.x}{\JournalTitle{\mnras},
  314, 33}

\bibitem[{{Betti} {et~al.}(2022{\natexlab{a}}){Betti}, {Follette},
  {Ward-Duong}, {Aoyama}, {Marleau}, {Bary}, {Robinson}, {Janson}, {Balmer},
  {Chauvin}, \& {Palma-Bifani}}]{betti22b}
{Betti}, S.~K., {Follette}, K.~B., {Ward-Duong}, K., {et~al.}
  2022{\natexlab{a}},
  \href{http://dx.doi.org/10.3847/2041-8213/ac85ef}{\JournalTitle{\apjl}, 935,
  L18}

\bibitem[{{Betti} {et~al.}(2022{\natexlab{b}}){Betti}, {Follette},
  {Ward-Duong}, {Aoyama}, {Marleau}, {Bary}, {Robinson}, {Janson}, {Balmer},
  {Chauvin}, \& {Palma-Bifani}}]{betti22c}
{Betti}, S.~K., {Follette}, K.~B., {Ward-Duong}, K., {et~al.}
  2022{\natexlab{b}},
  \href{http://dx.doi.org/10.3847/2041-8213/aca331}{\JournalTitle{\apjl}, 941,
  L20}

\bibitem[{{Betti} {et~al.}(2023){Betti}, {Follette}, {Ward-Duong}, {Peck},
  {Aoyama}, {Bary}, {Dacus}, {Edwards}, {Marleau}, {Mohamed}, {Palmo},
  {Plunkett}, {Robinson}, \& {Wang}}]{betti23}
{Betti}, S.~K., {Follette}, K.~B., {Ward-Duong}, K., {et~al.} 2023,
  \href{http://dx.doi.org/10.3847/1538-3881/ad06b8}{\JournalTitle{\aj}, 166,
  262}

\bibitem[{{B{\"o}ker} {et~al.}(2022){B{\"o}ker}, {Arribas}, {L{\"u}tzgendorf},
  {Alves de Oliveira}, {Beck}, {Birkmann}, {Bunker}, {Charlot}, {de Marchi},
  {Ferruit}, {Giardino}, {Jakobsen}, {Kumari}, {L{\'o}pez-Caniego}, {Maiolino},
  {Manjavacas}, {Marston}, {Moseley}, {Muzerolle}, {Ogle}, {Pirzkal},
  {Rauscher}, {Rawle}, {Rix}, {Sabbi}, {Sargent}, {Sirianni}, {te Plate},
  {Valenti}, {Willott}, \& {Zeidler}}]{boeker22}
{B{\"o}ker}, T., {Arribas}, S., {L{\"u}tzgendorf}, N., {et~al.} 2022,
  \href{http://dx.doi.org/10.1051/0004-6361/202142589}{\JournalTitle{\aap},
  661, A82}

\bibitem[{{Bonse} {et~al.}(2023){Bonse}, {Garvin}, {Gebhard}, {Dannert},
  {Cantalloube}, {Cugno}, {Absil}, {Hayoz}, {Milli}, {Kasper}, \&
  {Quanz}}]{bonse23}
{Bonse}, M.~J., {Garvin}, E.~O., {Gebhard}, T.~D., {et~al.} 2023,
  \href{http://dx.doi.org/10.3847/1538-3881/acc93c}{\JournalTitle{\aj}, 166,
  71}

\bibitem[{{Bowler} {et~al.}(2020){Bowler}, {Blunt}, \& {Nielsen}}]{bowler20}
{Bowler}, B.~P., {Blunt}, S.~C., \& {Nielsen}, E.~L. 2020,
  \href{http://dx.doi.org/10.3847/1538-3881/ab5b11}{\JournalTitle{\aj}, 159,
  63}

\bibitem[{{Brittain} {et~al.}(2020){Brittain}, {Najita}, {Dong}, \&
  {Zhu}}]{brittain20}
{Brittain}, S.~D., {Najita}, J.~R., {Dong}, R., \& {Zhu}, Z. 2020,
  \href{http://dx.doi.org/10.3847/1538-4357/ab8388}{\JournalTitle{\apj}, 895,
  48}

\bibitem[{{Calvet} \& {Gullbring}(1998)}]{calvetgull98}
{Calvet}, N., \& {Gullbring}, E. 1998,
  \href{http://dx.doi.org/10.1086/306527}{\JournalTitle{\apj}, 509, 802}

\bibitem[{{Chauvin} {et~al.}(2004){Chauvin}, {Lagrange}, {Dumas}, {Zuckerman},
  {Mouillet}, {Song}, {Beuzit}, \& {Lowrance}}]{chauvin04}
{Chauvin}, G., {Lagrange}, A.-M., {Dumas}, C., {et~al.} 2004,
  \href{http://dx.doi.org/10.1051/0004-6361:200400056}{\JournalTitle{\aap},
  425, L29}

\bibitem[{{Demars} {et~al.}(2023){Demars}, {Bonnefoy}, {Dougados}, {Aoyama},
  {Thanathibodee}, {Marleau}, {Tremblin}, {Delorme}, {Palma-Bifani}, {Petrus},
  {Bowler}, {Chauvin}, \& {Lagrange}}]{demars23}
{Demars}, D., {Bonnefoy}, M., {Dougados}, C., {et~al.} 2023,
  \href{http://dx.doi.org/10.1051/0004-6361/202346221}{\JournalTitle{\aap},
  676, A123}

\bibitem[{{Eriksson} {et~al.}(2020){Eriksson}, {Asensio Torres}, {Janson},
  {Aoyama}, {Marleau}, {Bonnefoy}, \& {Petrus}}]{eriksson20}
{Eriksson}, S.~C., {Asensio Torres}, R., {Janson}, M., {et~al.} 2020,
  \href{http://dx.doi.org/10.1051/0004-6361/202038131}{\JournalTitle{\aap},
  638, L6}

\bibitem[{{Erkal} {et~al.}(2022){Erkal}, {Manara}, {Schneider}, {Vincenzi},
  {Nisini}, {Coffey}, {Alcal{\'a}}, {Fedele}, \& {Antoniucci}}]{erkal22}
{Erkal}, J., {Manara}, C.~F., {Schneider}, P.~C., {et~al.} 2022,
  \href{http://dx.doi.org/10.1051/0004-6361/202244254}{\JournalTitle{\aap},
  666, A188}

\bibitem[{{Faherty} {et~al.}(2016){Faherty}, {Riedel}, {Cruz}, {Gagne},
  {Filippazzo}, {Lambrides}, {Fica}, {Weinberger}, {Thorstensen}, {Tinney},
  {Baldassare}, {Lemonier}, \& {Rice}}]{faherty16}
{Faherty}, J.~K., {Riedel}, A.~R., {Cruz}, K.~L., {et~al.} 2016,
  \href{http://dx.doi.org/10.3847/0067-0049/225/1/10}{\JournalTitle{\apjs},
  225, 10}

\bibitem[{{Fischer} {et~al.}(2008){Fischer}, {Kwan}, {Edwards}, \&
  {Hillenbrand}}]{fischer08}
{Fischer}, W., {Kwan}, J., {Edwards}, S., \& {Hillenbrand}, L. 2008,
  \href{http://dx.doi.org/10.1086/591902}{\JournalTitle{\apj}, 687, 1117}

\bibitem[{{Gaia Collaboration} {et~al.}(2021){Gaia Collaboration}, {Brown},
  {Vallenari}, {Prusti}, {de Bruijne}, {Babusiaux}, {Biermann}, {Creevey},
  {Evans}, {Eyer}, {Hutton}, {Jansen}, {Jordi}, {Klioner}, {Lammers},
  {Lindegren}, {Luri}, {Mignard}, {Panem}, {Pourbaix}, {Randich}, {Sartoretti},
  {Soubiran}, {Walton}, {Arenou}, {Bailer-Jones}, {Bastian}, {Cropper},
  {Drimmel}, {Katz}, {Lattanzi}, {van Leeuwen}, {Bakker}, {Cacciari},
  {Casta{\~n}eda}, {De Angeli}, {Ducourant}, {Fabricius}, {Fouesneau},
  {Fr{\'e}mat}, {Guerra}, {Guerrier}, {Guiraud}, {Jean-Antoine Piccolo},
  {Masana}, {Messineo}, {Mowlavi}, {Nicolas}, {Nienartowicz}, {Pailler},
  {Panuzzo}, {Riclet}, {Roux}, {Seabroke}, {Sordo}, {Tanga}, {Th{\'e}venin},
  {Gracia-Abril}, {Portell}, {Teyssier}, {Altmann}, {Andrae}, {Bellas-Velidis},
  {Benson}, {Berthier}, {Blomme}, {Brugaletta}, {Burgess}, {Busso}, {Carry},
  {Cellino}, {Cheek}, {Clementini}, {Damerdji}, {Davidson}, {Delchambre},
  {Dell'Oro}, {Fern{\'a}ndez-Hern{\'a}ndez}, {Galluccio}, {Garc{\'\i}a-Lario},
  {Garcia-Reinaldos}, {Gonz{\'a}lez-N{\'u}{\~n}ez}, {Gosset}, {Haigron},
  {Halbwachs}, {Hambly}, {Harrison}, {Hatzidimitriou}, {Heiter},
  {Hern{\'a}ndez}, {Hestroffer}, {Hodgkin}, {Holl}, {Jan{\ss}en}, {Jevardat de
  Fombelle}, {Jordan}, {Krone-Martins}, {Lanzafame}, {L{\"o}ffler}, {Lorca},
  {Manteiga}, {Marchal}, {Marrese}, {Moitinho}, {Mora}, {Muinonen}, {Osborne},
  {Pancino}, {Pauwels}, {Petit}, {Recio-Blanco}, {Richards}, {Riello},
  {Rimoldini}, {Robin}, {Roegiers}, {Rybizki}, {Sarro}, {Siopis}, {Smith},
  {Sozzetti}, {Ulla}, {Utrilla}, {van Leeuwen}, {van Reeven}, {Abbas}, {Abreu
  Aramburu}, {Accart}, {Aerts}, {Aguado}, {Ajaj}, {Altavilla}, {{\'A}lvarez},
  {{\'A}lvarez Cid-Fuentes}, {Alves}, {Anderson}, {Anglada Varela}, {Antoja},
  {Audard}, {Baines}, {Baker}, {Balaguer-N{\'u}{\~n}ez}, {Balbinot}, {Balog},
  {Barache}, {Barbato}, {Barros}, {Barstow}, {Bartolom{\'e}}, {Bassilana},
  {Bauchet}, {Baudesson-Stella}, {Becciani}, {Bellazzini}, {Bernet}, {Bertone},
  {Bianchi}, {Blanco-Cuaresma}, {Boch}, {Bombrun}, {Bossini}, {Bouquillon},
  {Bragaglia}, {Bramante}, {Breedt}, {Bressan}, {Brouillet}, {Bucciarelli},
  {Burlacu}, {Busonero}, {Butkevich}, {Buzzi}, {Caffau}, {Cancelliere},
  {C{\'a}novas}, {Cantat-Gaudin}, {Carballo}, {Carlucci}, {Carnerero},
  {Carrasco}, {Casamiquela}, {Castellani}, {Castro-Ginard}, {Castro Sampol},
  {Chaoul}, {Charlot}, {Chemin}, {Chiavassa}, {Cioni}, {Comoretto}, {Cooper},
  {Cornez}, {Cowell}, {Crifo}, {Crosta}, {Crowley}, {Dafonte}, {Dapergolas},
  {David}, {David}, {de Laverny}, {De Luise}, {De March}, {De Ridder}, {de
  Souza}, {de Teodoro}, {de Torres}, {del Peloso}, {del Pozo}, {Delbo},
  {Delgado}, {Delgado}, {Delisle}, {Di Matteo}, {Diakite}, {Diener},
  {Distefano}, {Dolding}, {Eappachen}, {Edvardsson}, {Enke}, {Esquej}, {Fabre},
  {Fabrizio}, {Faigler}, {Fedorets}, {Fernique}, {Fienga}, {Figueras},
  {Fouron}, {Fragkoudi}, {Fraile}, {Franke}, {Gai}, {Garabato},
  {Garcia-Gutierrez}, {Garc{\'\i}a-Torres}, {Garofalo}, {Gavras}, {Gerlach},
  {Geyer}, {Giacobbe}, {Gilmore}, {Girona}, {Giuffrida}, {Gomel}, {Gomez},
  {Gonzalez-Santamaria}, {Gonz{\'a}lez-Vidal}, {Granvik},
  {Guti{\'e}rrez-S{\'a}nchez}, {Guy}, {Hauser}, {Haywood}, {Helmi}, {Hidalgo},
  {Hilger}, {H{\l}adczuk}, {Hobbs}, {Holland}, {Huckle}, {Jasniewicz},
  {Jonker}, {Juaristi Campillo}, {Julbe}, {Karbevska}, {Kervella}, {Khanna},
  {Kochoska}, {Kontizas}, {Kordopatis}, {Korn}, {Kostrzewa-Rutkowska},
  {Kruszy{\'n}ska}, {Lambert}, {Lanza}, {Lasne}, {Le Campion}, {Le Fustec},
  {Lebreton}, {Lebzelter}, {Leccia}, {Leclerc}, {Lecoeur-Taibi}, {Liao},
  {Licata}, {Lindstr{\o}m}, {Lister}, {Livanou}, {Lobel}, {Madrero Pardo},
  {Managau}, {Mann}, {Marchant}, {Marconi}, {Marcos Santos}, {Marinoni},
  {Marocco}, {Marshall}, {Martin Polo}, {Mart{\'\i}n-Fleitas}, {Masip},
  {Massari}, {Mastrobuono-Battisti}, {Mazeh}, {McMillan}, {Messina},
  {Michalik}, {Millar}, {Mints}, {Molina}, {Molinaro}, {Moln{\'a}r},
  {Montegriffo}, {Mor}, {Morbidelli}, {Morel}, {Morris}, {Mulone}, {Munoz},
  {Muraveva}, {Murphy}, {Musella}, {Noval}, {Ord{\'e}novic}, {Orr{\`u}},
  {Osinde}, {Pagani}, {Pagano}, {Palaversa}, {Palicio}, {Panahi}, {Pawlak},
  {Pe{\~n}alosa Esteller}, {Penttil{\"a}}, {Piersimoni}, {Pineau}, {Plachy},
  {Plum}, {Poggio}, {Poretti}, {Poujoulet}, {Pr{\v{s}}a}, {Pulone}, {Racero},
  {Ragaini}, {Rainer}, {Raiteri}, {Rambaux}, {Ramos}, {Ramos-Lerate}, {Re
  Fiorentin}, {Regibo}, {Reyl{\'e}}, {Ripepi}, {Riva}, {Rixon}, {Robichon},
  {Robin}, {Roelens}, {Rohrbasser}, {Romero-G{\'o}mez}, {Rowell}, {Royer},
  {Rybicki}, {Sadowski}, {Sagrist{\`a} Sell{\'e}s}, {Sahlmann}, {Salgado},
  {Salguero}, {Samaras}, {Sanchez Gimenez}, {Sanna}, {Santove{\~n}a},
  {Sarasso}, {Schultheis}, {Sciacca}, {Segol}, {Segovia}, {S{\'e}gransan},
  {Semeux}, {Shahaf}, {Siddiqui}, {Siebert}, {Siltala}, {Slezak}, {Smart},
  {Solano}, {Solitro}, {Souami}, {Souchay}, {Spagna}, {Spoto}, {Steele},
  {Steidelm{\"u}ller}, {Stephenson}, {S{\"u}veges}, {Szabados}, {Szegedi-Elek},
  {Taris}, {Tauran}, {Taylor}, {Teixeira}, {Thuillot}, {Tonello}, {Torra},
  {Torra}, {Turon}, {Unger}, {Vaillant}, {van Dillen}, {Vanel}, {Vecchiato},
  {Viala}, {Vicente}, {Voutsinas}, {Weiler}, {Wevers}, {Wyrzykowski}, {Yoldas},
  {Yvard}, {Zhao}, {Zorec}, {Zucker}, {Zurbach}, \& {Zwitter}}]{gEDR3}
{Gaia Collaboration}, {Brown}, A.~G.~A., {Vallenari}, A., {et~al.} 2021,
  \href{http://dx.doi.org/10.1051/0004-6361/202039657}{\JournalTitle{\aap},
  649, A1}

\bibitem[{{Gangi} {et~al.}(2022){Gangi}, {Antoniucci}, {Biazzo}, {Frasca},
  {Nisini}, {Alcal{\'a}}, {Giannini}, {Manara}, {Giunta}, {Harutyunyan},
  {Munari}, \& {Vitali}}]{gangi22}
{Gangi}, M., {Antoniucci}, S., {Biazzo}, K., {et~al.} 2022,
  \href{http://dx.doi.org/10.1051/0004-6361/202244042}{\JournalTitle{\aap},
  667, A124}

\bibitem[{{Gardner} {et~al.}(2023){Gardner}, {Mather}, {Abbott}, {Abell},
  {Abernathy}, {Abney}, {Abraham}, {Abraham}, {Abul-Huda}, {Acton}, {Adams},
  {Adams}, {Adler}, {Adriaensen}, {Aguilar}, {Ahmed}, {Ahmed}, {Ahmed},
  {Albat}, {Albert}, {Alberts}, {Aldridge}, {Allen}, {Allen}, {Altenburg},
  {Altunc}, {Alvarez}, {{\'A}lvarez-M{\'a}rquez}, {Alves de Oliveira},
  {Ambrose}, {Anandakrishnan}, {Andersen}, {Anderson}, {Anderson}, {Anderson},
  {Anderson}, {Aprea}, {Archer}, {Arenberg}, {Argyriou}, {Arribas}, {Artigau},
  {Arvai}, {Atcheson}, {Atkinson}, {Averbukh}, {Aymergen}, {Bacinski},
  {Baggett}, {Bagnasco}, {Baker}, {Balzano}, {Banks}, {Baran}, {Barker},
  {Barrett}, {Barringer}, {Barto}, {Bast}, {Baudoz}, {Baum}, {Beatty},
  {Beaulieu}, {Bechtold}, {Beck}, {Beddard}, {Beichman}, {Bellagama}, {Bely},
  {Berger}, {Bergeron}, {Bernier}, {Bertch}, {Beskow}, {Betz}, {Biagetti},
  {Birkmann}, {Bjorklund}, {Blackwood}, {Blazek}, {Blossfeld}, {Bluth},
  {Boccaletti}, {Boegner}, {Bohlin}, {Boia}, {B{\"o}ker}, {Bonaventura},
  {Bond}, {Bosley}, {Boucarut}, {Bouchet}, {Bouwman}, {Bower}, {Bowers},
  {Bowers}, {Boyce}, {Boyer}, {Boyer}, {Boyer}, {Boyer}, {Bradley}, {Brady},
  {Brandl}, {Brannen}, {Breda}, {Bremmer}, {Brennan}, {Bresnahan}, {Bright},
  {Broiles}, {Bromenschenkel}, {Brooks}, {Brooks}, {Brown}, {Brown}, {Brown},
  {Bruce}, {Bryson}, {Bujanda}, {Bullock}, {Bunker}, {Bureo}, {Burt}, {Bush},
  {Bushouse}, {Bussman}, {Cabaud}, {Cale}, {Calhoon}, {Calvani}, {Canipe},
  {Caputo}, {Cara}, {Carey}, {Case}, {Cesari}, {Cetorelli}, {Chance},
  {Chandler}, {Chaney}, {Chapman}, {Charlot}, {Chayer}, {Cheezum}, {Chen},
  {Chen}, {Cherinka}, {Chichester}, {Chilton}, {Chittiraibalan}, {Clampin},
  {Clark}, {Clark}, {Clark}, {Claybrooks}, {Cleveland}, {Cohen}, {Cohen},
  {Col{\'o}n}, {Coleman}, {Colina}, {Comber}, {Comeau}, {Comer}, {Conde Reis},
  {Connolly}, {Conroy}, {Contos}, {Contreras}, {Cook}, {Cooper}, {Cooper},
  {Correia}, {Correnti}, {Cossou}, {Costanza}, {Coulais}, {Cox}, {Coyle},
  {Cracraft}, {Crew}, {Curtis}, {Cusveller}, {Da Costa Maciel}, {Dailey},
  {Daugeron}, {Davidson}, {Davies}, {Davis}, {Davis}, {Day}, {de Chambure}, {de
  Jong}, {De Marchi}, {Dean}, {Decker}, {Delisa}, {Dell}, {Dellagatta},
  {Dembinska}, {Demosthenes}, {Dencheva}, {Deneu}, {DePriest}, {Deschenes},
  {Dethienne}, {Detre}, {Diaz}, {Dicken}, {DiFelice}, {Dillman}, {Disharoon},
  {Dixon}, {Doggett}, {Dominguez}, {Donaldson}, {Doria-Warner}, {Santos},
  {Doty}, {Douglas}, {Doyon}, {Dressler}, {Driggers}, {Driggers}, {Dunn},
  {DuPrie}, {Dupuis}, {Durning}, {Dutta}, {Earl}, {Eccleston}, {Ecobichon},
  {Egami}, {Ehrenwinkler}, {Eisenhamer}, {Eisenhower}, {Eisenstein}, {El
  Hamel}, {Elie}, {Elliott}, {Elliott}, {Engesser}, {Espinoza}, {Etienne},
  {Etxaluze}, {Evans}, {Fabreguettes}, {Falcolini}, {Falini}, {Fatig},
  {Feeney}, {Feinberg}, {Fels}, {Ferdous}, {Ferguson}, {Ferrarese}, {Ferreira},
  {Ferruit}, {Ferry}, {Filippazzo}, {Firre}, {Fix}, {Flagey}, {Flanagan},
  {Fleming}, {Florian}, {Flynn}, {Foiadelli}, {Fontaine}, {Fontanella},
  {Forshay}, {Fortner}, {Fox}, {Framarini}, {Francisco}, {Franck}, {Franx},
  {Franz}, {Friedman}, {Friend}, {Frost}, {Fu}, {Fullerton}, {Gaillard},
  {Galkin}, {Gallagher}, {Galyer}, {Garc{\'\i}a Mar{\'\i}n}, {Gardner},
  {Garland}, {Garrett}, {Gasman}, {G{\'a}sp{\'a}r}, {Gastaud}, {Gaudreau},
  {Gauthier}, {Geers}, {Geithner}, {Gennaro}, {Gerber}, {Gereau}, {Giampaoli},
  {Giardino}, {Gibbons}, {Gilbert}, {Gilman}, {Girard}, {Giuliano}, {Gkountis},
  {Glasse}, {Glassmire}, {Glauser}, {Glazer}, {Goldberg}, {Golimowski},
  {Gonzaga}, {Gordon}, {Gordon}, {Goudfrooij}, {Gough}, {Graham}, {Grau},
  {Green}, {Greene}, {Greene}, {Greenfield}, {Greenhouse}, {Greve}, {Greville},
  {Grimaldi}, {Groe}, {Groebner}, {Grumm}, {Grundy}, {G{\"u}del}, {Guillard},
  {Guldalian}, {Gunn}, {Gurule}, {Gutman}, {Guy}, {Guyot}, {Hack}, {Haderlein},
  {Hagan}, {Hagedorn}, {Hainline}, {Haley}, {Hami}, {Hamilton}, {Hammann},
  {Hammel}, {Hanley}, {Hansen}, {Hardy}, {Harnisch}, {Harr}, {Harris}, {Hart},
  {Hartig}, {Hasan}, {Hashim}, {Hashimoto}, {Haskins}, {Hawkins}, {Hayden},
  {Hayden}, {Healy}, {Hecht}, {Heeg}, {Hejal}, {Helm}, {Hengemihle}, {Henning},
  {Henry}, {Henry}, {Henshaw}, {Hernandez}, {Herrington}, {Heske}, {Hesman},
  {Hickey}, {Hilbert}, {Hines}, {Hinz}, {Hirsch}, {Hitcho}, {Hodapp}, {Hodge},
  {Hoffman}, {Holfeltz}, {Holler}, {Hoppa}, {Horner}, {Howard}, {Howard},
  {Huber}, {Hunkeler}, {Hunter}, {Hunter}, {Hurd}, {Hurst}, {Hutchings},
  {Hylan}, {Ignat}, {Illingworth}, {Irish}, {Isaacs}, {Jackson}, {Jaffe},
  {Jahic}, {Jahromi}, {Jakobsen}, {James}, {James}, {James}, {Jamieson},
  {Jandra}, {Jayawardhana}, {Jedrzejewski}, {Jeffers}, {Jensen}, {Joanne},
  {Johns}, {Johnson}, {Johnson}, {Johnson}, {Johnson}, {Johnson}, {Johnson},
  {Johnstone}, {Jollet}, {Jones}, {Jones}, {Jones}, {Jones}, {Jones}, {Jordan},
  {Jordan}, {Jue}, {Jurkowski}, {Justis}, {Justtanont}, {Kaleida}, {Kalirai},
  {Kalmanson}, {Kaltenegger}, {Kammerer}, {Kan}, {Kanarek}, {Kao}, {Karakla},
  {Karl}, {Kassin}, {Kauffman}, {Kavanagh}, {Kelley}, {Kelly}, {Kendrew},
  {Kennedy}, {Kenny}, {Keski-Kuha}, {Keyes}, {Khan}, {Kidwell}, {Kimble},
  {King}, {King}, {Kinzel}, {Kirk}, {Kirkpatrick}, {Klaassen}, {Klingemann},
  {Klintworth}, {Knapp}, {Knight}, {Knollenberg}, {Knutsen}, {Koehler},
  {Koekemoer}, {Kofler}, {Kontson}, {Kovacs}, {Kozhurina-Platais}, {Krause},
  {Kriss}, {Krist}, {Kristoffersen}, {Krogel}, {Krueger}, {Kulp}, {Kumari},
  {Kwan}, {Kyprianou}, {Labador}, {Labiano}, {Lafreni{\`e}re}, {Lagage},
  {Laidler}, {Laine}, {Laird}, {Lajoie}, {Lallo}, {Lam}, {LaMassa}, {Lambros},
  {Lampenfield}, {Lander}, {Langston}, {Larson}, {Larson}, {LaVerghetta},
  {Law}, {Lawrence}, {Lee}, {Lee}, {Lee}, {Leisenring}, {Leveille}, {Levenson},
  {Levi}, {Levine}, {Lewis}, {Lewis}, {Lewis}, {Libralato}, {Lidon},
  {Liebrecht}, {Lightsey}, {Lilly}, {Lim}, {Lim}, {Ling}, {Link}, {Link},
  {Lipinski}, {Liu}, {Lo}, {Lobmeyer}, {Logue}, {Long}, {Long}, {Long}, {Long},
  {L{\'o}pez-Caniego}, {Lotz}, {Love-Pruitt}, {Lubskiy}, {Luers}, {Luetgens},
  {Luevano}, {Lui}, {Lund}, {Lundquist}, {Lunine}, {L{\"u}tzgendorf}, {Lynch},
  {MacDonald}, {MacDonald}, {Macias}, {Macklis}, {Maghami}, {Maharaja},
  {Maiolino}, {Makrygiannis}, {Malla}, {Malumuth}, {Manjavacas}, {Marini},
  {Marrione}, {Marston}, {Martel}, {Martin}, {Martin}, {Martinez}, {Maschmann},
  {Masci}, {Masetti}, {Maszkiewicz}, {Matthews}, {Matuskey}, {McBrayer},
  {McCarthy}, {McCaughrean}, {McClare}, {McClare}, {McCloskey}, {McClurg},
  {McCoy}, {McElwain}, {McGregor}, {McGuffey}, {McKay}, {McKenzie}, {McLean},
  {McMaster}, {McNeil}, {De Meester}, {Mehalick}, {Meixner}, {Mel{\'e}ndez},
  {Menzel}, {Menzel}, {Merz}, {Mesterharm}, {Meyer}, {Meyett}, {Meza},
  {Midwinter}, {Milam}, {Miller}, {Miller}, {Miskey}, {Misselt}, {Mitchell},
  {Mohan}, {Montoya}, {Moran}, {Morishita}, {Moro-Mart{\'\i}n}, {Morrison},
  {Morrison}, {Morse}, {Moschos}, {Moseley}, {Mosier}, {Mosner}, {Mountain},
  {Muckenthaler}, {Mueller}, {Mueller}, {Muhiem}, {M{\"u}hlmann}, {Mullally},
  {Mullen}, {Munger}, {Murphy}, {Murray}, {Muzerolle}, {Mycroft}, {Myers},
  {Myers}, {Myers}, {Myers}, {Myrick}, {Nagle}, {Nayak}, {Naylor}, {Neff},
  {Nelan}, {Nella}, {Nguyen}, {Nguyen}, {Nickson}, {Nidhiry}, {Niedner},
  {Nieto-Santisteban}, {Nikolov}, {Nishisaka}, {Noriega-Crespo}, {Nota},
  {O'Mara}, {Oboryshko}, {O'Brien}, {Ochs}, {Offenberg}, {Ogle}, {Ohl},
  {Olmsted}, {Osborne}, {O'Shaughnessy}, {{\"O}stlin}, {O'Sullivan}, {Otor},
  {Ottens}, {Ouellette}, {Outlaw}, {Owens}, {Pacifici}, {Page}, {Paranilam},
  {Park}, {Parrish}, {Paschal}, {Patapis}, {Patel}, {Patrick}, {Pattishall},
  {Paul}, {Paul}, {Pauly}, {Pavlovsky}, {Pe{\~n}a-Guerrero}, {Pedder}, {Peek},
  {Pelham}, {Penanen}, {Perriello}, {Perrin}, {Perrine}, {Perrygo}, {Peslier},
  {Petach}, {Peterson}, {Pfarr}, {Pierson}, {Pietraszkiewicz}, {Pilchen},
  {Pipher}, {Pirzkal}, {Pitman}, {Player}, {Plesha}, {Plitzke}, {Pohner},
  {Poletis}, {Pollizzi}, {Polster}, {Pontius}, {Pontoppidan}, {Porges},
  {Potter}, {Prescott}, {Proffitt}, {Pueyo}, {Quispe Neira}, {Radich}, {Rager},
  {Rameau}, {Ramey}, {Ramos Alarcon}, {Rampini}, {Rapp}, {Rashford},
  {Rauscher}, {Ravindranath}, {Rawle}, {Rawlings}, {Ray}, {Regan}, {Rehm},
  {Rehm}, {Reid}, {Reis}, {Renk}, {Reoch}, {Ressler}, {Rest}, {Reynolds},
  {Richon}, {Richon}, {Ridgaway}, {Riedel}, {Rieke}, {Rieke}, {Rifelli},
  {Rigby}, {Riggs}, {Ringel}, {Ritchie}, {Rix}, {Robberto}, {Robinson},
  {Robinson}, {Robinson}, {Rock}, {Rodriguez}, {Rodr{\'\i}guez del Pino},
  {Roellig}, {Rohrbach}, {Roman}, {Romelfanger}, {Romo}, {Rosales}, {Rose},
  {Roteliuk}, {Roth}, {Rothwell}, {Rouzaud}, {Rowe}, {Rowlands}, {Roy},
  {Royer}, {Rui}, {Rumler}, {Rumpl}, {Russ}, {Ryan}, {Ryan}, {Saad}, {Sabata},
  {Sabatino}, {Sabbi}, {Sabelhaus}, {Sabia}, {Sahu}, {Saif}, {Salvignol},
  {Samara-Ratna}, {Samuelson}, {Sanders}, {Sappington}, {Sargent}, {Sauer},
  {Savadkin}, {Sawicki}, {Schappell}, {Scheffer}, {Scheithauer}, {Scherer},
  {Schiff}, {Schlawin}, {Schmeitzky}, {Schmitz}, {Schmude}, {Schneider},
  {Schreiber}, {Schroeven-Deceuninck}, {Schultz}, {Schwab}, {Schwartz},
  {Scoccimarro}, {Scott}, {Scott}, {Seaton}, {Seely}, {Seery}, {Seidleck},
  {Sembach}, {Shanahan}, {Shaughnessy}, {Shaw}, {Shay}, {Sheehan}, {Sheth},
  {Shih}, {Shivaei}, {Siegel}, {Sienkiewicz}, {Simmons}, {Simon}, {Sirianni},
  {Sivaramakrishnan}, {Slade}, {Sloan}, {Slocum}, {Slowinski}, {Smith},
  {Smith}, {Smith}, {Smith}, {Smith}, {Smith}, {Smolik}, {Soderblom}, {Sohn},
  {Sokol}, {Sonneborn}, {Sontag}, {Sooy}, {Soummer}, {Southwood}, {Spain},
  {Sparmo}, {Speer}, {Spencer}, {Sprofera}, {Stallcup}, {Stanley},
  {Stansberry}, {Stark}, {Starr}, {Stassi}, {Steck}, {Steeley}, {Stephens},
  {Stephenson}, {Stewart}, {Stiavelli}, {}, {Strada}, {Straughn}, {Streetman},
  {Strickland}, {Strobele}, {Stuhlinger}, {Stys}, {Such}, {Sukhatme},
  {Sullivan}, {Sullivan}, {Sumner}, {Sun}, {Sunnquist}, {Swade}, {Swam},
  {Swenton}, {Swoish}, {Tam Litten}, {Tamas}, {Tao}, {Taylor}, {Taylor}, {te
  Plate}, {Van Tea}, {Teague}, {Telfer}, {Temim}, {Texter}, {Thatte},
  {Thompson}, {Thompson}, {Thomson}, {Thronson}, {Tierney}, {Tikkanen},
  {Tinnin}, {Tippet}, {Todd}, {Tran}, {Trauger}, {Trejo}, {Vinh Truong},
  {Tsukamoto}, {Tufail}, {Tumlinson}, {Tustain}, {Tyra}, {Ubeda}, {Underwood},
  {Uzzo}, {Vaclavik}, {Valenduc}, {Valenti}, {Van Campen}, {van de Wetering},
  {Van Der Marel}, {van Haarlem}, {Vandenbussche}, {van Dishoeck},
  {Vanterpool}, {Vernoy}, {Vila Costas}, {Volk}, {Voorzaat}, {Voyton}, {Vydra},
  {Waddy}, {Waelkens}, {Wahlgren}, {Walker}, {Wander}, {Warfield}, {Warner},
  {Wasiak}, {Wasiak}, {Wehner}, {Weiler}, {Weilert}, {Weiss}, {Wells}, {Welty},
  {Wheate}, {Wheeler}, {White}, {Whitehouse}, {Whiteleather}, {Whitman},
  {Williams}, {Willmer}, {Willott}, {Willoughby}, {Wilson}, {Wilson}, {Wilson},
  {Windhorst}, {Wislowski}, {Wolfe}, {Wolfe}, {Wolff}, {Wondel}, {Woo},
  {Woods}, {Worden}, {Workman}, {Wright}, {Wu}, {Wu}, {Wun}, {Wymer},
  {Yadetie}, {Yan}, {Yang}, {Yates}, {Yeager}, {Yerger}, {Young}, {Young},
  {Yu}, {Yu}, {Zak}, {Zeidler}, {Zepp}, {Zhou}, {Zincke}, {Zonak}, \&
  {Zondag}}]{gardner23}
{Gardner}, J.~P., {Mather}, J.~C., {Abbott}, R., {et~al.} 2023,
  \href{http://dx.doi.org/10.1088/1538-3873/acd1b5}{\JournalTitle{\pasp}, 135,
  068001}

\bibitem[{{Gizis}(2002)}]{gizis02}
{Gizis}, J.~E. 2002,
  \href{http://dx.doi.org/10.1086/341259}{\JournalTitle{\apj}, 575, 484}

\bibitem[{{Haffert} {et~al.}(2019){Haffert}, {Bohn}, {de Boer}, {Snellen},
  {Brinchmann}, {Girard}, {Keller}, \& {Bacon}}]{Haffert+2019}
{Haffert}, S.~Y., {Bohn}, A.~J., {de Boer}, J., {et~al.} 2019,
  \href{http://dx.doi.org/10.1038/s41550-019-0780-5}{\JournalTitle{\natas}, 3,
  749}

\bibitem[{{Hartmann} {et~al.}(2016){Hartmann}, {Herczeg}, \&
  {Calvet}}]{hartmann16}
{Hartmann}, L., {Herczeg}, G., \& {Calvet}, N. 2016,
  \href{http://dx.doi.org/10.1146/annurev-astro-081915-023347}{\JournalTitle{\araa},
  54, 135}

\bibitem[{{Hasegawa} {et~al.}(2024){Hasegawa}, {Uyama}, {Hashimoto}, {Aoyama},
  {Deo}, {Guyon}, {Lozi}, {Norris}, {Tamura}, \& {Vievard}}]{hasegawa24}
{Hasegawa}, Y., {Uyama}, T., {Hashimoto}, J., {et~al.} 2024,
  \href{http://dx.doi.org/10.3847/1538-3881/ad1cec}{\JournalTitle{\aj}, 167,
  105}

\bibitem[{{Herczeg} {et~al.}(2009){Herczeg}, {Cruz}, \&
  {Hillenbrand}}]{herczeg09}
{Herczeg}, G.~J., {Cruz}, K.~L., \& {Hillenbrand}, L.~A. 2009,
  \href{http://dx.doi.org/10.1088/0004-637X/696/2/1589}{\JournalTitle{\apj},
  696, 1589}

\bibitem[{{Herczeg} {et~al.}(2004){Herczeg}, {Wood}, {Linsky}, {Valenti}, \&
  {Johns-Krull}}]{herczeg04}
{Herczeg}, G.~J., {Wood}, B.~E., {Linsky}, J.~L., {Valenti}, J.~A., \&
  {Johns-Krull}, C.~M. 2004,
  \href{http://dx.doi.org/10.1086/383340}{\JournalTitle{\apj}, 607, 369}

\bibitem[{{Herczeg} {et~al.}(2023){Herczeg}, {Chen}, {Donati}, {Dupree},
  {Walter}, {Hillenbrand}, {Johns-Krull}, {Manara}, {G{\"u}nther}, {Fang},
  {Schneider}, {Valenti}, {Alencar}, {Venuti}, {Alcal{\'a}}, {Frasca},
  {Arulanantham}, {Linsky}, {Bouvier}, {Brickhouse}, {Calvet}, {Espaillat},
  {Campbell-White}, {Carpenter}, {Chang}, {Cruz}, {Dahm}, {Eisl{\"o}ffel},
  {Edwards}, {Fischer}, {Guo}, {Henning}, {Ji}, {Jose}, {Kastner}, {Launhardt},
  {Principe}, {Robinson}, {Serna}, {Siwak}, {Sterzik}, \&
  {Takasao}}]{herczeg23}
{Herczeg}, G.~J., {Chen}, Y., {Donati}, J.-F., {et~al.} 2023,
  \href{http://dx.doi.org/10.3847/1538-4357/acf468}{\JournalTitle{\apj}, 956,
  102}

\bibitem[{{Jakobsen} {et~al.}(2022){Jakobsen}, {Ferruit}, {Alves de Oliveira},
  {Arribas}, {Bagnasco}, {Barho}, {Beck}, {Birkmann}, {B{\"o}ker}, {Bunker},
  {Charlot}, {de Jong}, {de Marchi}, {Ehrenwinkler}, {Falcolini}, {Fels},
  {Franx}, {Franz}, {Funke}, {Giardino}, {Gnata}, {Holota}, {Honnen}, {Jensen},
  {Jentsch}, {Johnson}, {Jollet}, {Karl}, {Kling}, {K{\"o}hler}, {Kolm},
  {Kumari}, {Lander}, {Lemke}, {L{\'o}pez-Caniego}, {L{\"u}tzgendorf},
  {Maiolino}, {Manjavacas}, {Marston}, {Maschmann}, {Maurer}, {Messerschmidt},
  {Moseley}, {Mosner}, {Mott}, {Muzerolle}, {Pirzkal}, {Pittet}, {Plitzke},
  {Posselt}, {Rapp}, {Rauscher}, {Rawle}, {Rix}, {R{\"o}del}, {Rumler},
  {Sabbi}, {Salvignol}, {Schmid}, {Sirianni}, {Smith}, {Strada}, {te Plate},
  {Valenti}, {Wettemann}, {Wiehe}, {Wiesmayer}, {Willott}, {Wright}, {Zeidler},
  \& {Zincke}}]{jakobsen22}
{Jakobsen}, P., {Ferruit}, P., {Alves de Oliveira}, C., {et~al.} 2022,
  \href{http://dx.doi.org/10.1051/0004-6361/202142663}{\JournalTitle{\aap},
  661, A80}

\bibitem[{{Komarova} \& {Fischer}(2020)}]{Komarova+Fischer2020}
{Komarova}, O., \& {Fischer}, W.~J. 2020,
  \href{http://dx.doi.org/10.3847/2515-5172/ab67bb}{\JournalTitle{RNAAS}, 4, 6}

\bibitem[{{Kwan} {et~al.}(2007){Kwan}, {Edwards}, \& {Fischer}}]{kwan07}
{Kwan}, J., {Edwards}, S., \& {Fischer}, W. 2007,
  \href{http://dx.doi.org/10.1086/511057}{\JournalTitle{\apj}, 657, 897}

\bibitem[{{Lai} \& {Mu\~{n}oz}(2023)}]{lai23}
{Lai}, D., \& {Mu\~{n}oz}, D.~J. 2023,
  \href{http://dx.doi.org/10.1146/annurev-astro-052622-022933}{\JournalTitle{\araa},
  61, 517}

\bibitem[{{Lodato} {et~al.}(2005){Lodato}, {Delgado-Donate}, \&
  {Clarke}}]{lodato05}
{Lodato}, G., {Delgado-Donate}, E., \& {Clarke}, C.~J. 2005,
  \href{http://dx.doi.org/10.1111/j.1745-3933.2005.00112.x}{\JournalTitle{\mnras},
  364, L91}

\bibitem[{{Looper} {et~al.}(2010){Looper}, {Bochanski}, {Burgasser}, {Mohanty},
  {Mamajek}, {Faherty}, {West}, \& {Pitts}}]{looper10b}
{Looper}, D.~L., {Bochanski}, J.~J., {Burgasser}, A.~J., {et~al.} 2010,
  \href{http://dx.doi.org/10.1088/0004-6256/140/5/1486}{\JournalTitle{\aj},
  140, 1486}

\bibitem[{{Luhman}(2023)}]{luhman23b}
{Luhman}, K.~L. 2023,
  \href{http://dx.doi.org/10.3847/1538-3881/accf19}{\JournalTitle{\aj}, 165,
  269}

\bibitem[{{Luhman} {et~al.}(2007){Luhman}, {Adame}, {D'Alessio}, {Calvet},
  {McLeod}, {Bohac}, {Forrest}, {Hartmann}, {Sargent}, \& {Watson}}]{luhman07d}
{Luhman}, K.~L., {Adame}, L., {D'Alessio}, P., {et~al.} 2007,
  \href{http://dx.doi.org/10.1086/520712}{\JournalTitle{\apj}, 666, 1219}

\bibitem[{{Luhman} {et~al.}(2023){Luhman}, {Tremblin}, {Birkmann},
  {Manjavacas}, {Valenti}, {Alves de Oliveira}, {Beck}, {Giardino},
  {L{\"u}tzgendorf}, {Rauscher}, \& {Sirianni}}]{luhman23c}
{Luhman}, K.~L., {Tremblin}, P., {Birkmann}, S.~M., {et~al.} 2023,
  \href{http://dx.doi.org/10.3847/2041-8213/acd635}{\JournalTitle{\apjl}, 949,
  L36}

\bibitem[{{Lyons}(2013)}]{lyons13}
{Lyons}, L. 2013,
  \href{http://dx.doi.org/10.48550/arXiv.1310.1284}{\JournalTitle{arXiv
  e-prints}, arXiv:1310.1284}

\bibitem[{{Manara} {et~al.}(2017){Manara}, {Frasca}, {Alcal{\'a}}, {Natta},
  {Stelzer}, \& {Testi}}]{manara17b}
{Manara}, C.~F., {Frasca}, A., {Alcal{\'a}}, J.~M., {et~al.} 2017,
  \href{http://dx.doi.org/10.1051/0004-6361/201730807}{\JournalTitle{\aap},
  605, A86}

\bibitem[{{Manara} {et~al.}(2013){Manara}, {Testi}, {Rigliaco}, {Alcal{\'a}},
  {Natta}, {Stelzer}, {Biazzo}, {Covino}, {Covino}, {Cupani}, {D'Elia}, \&
  {Randich}}]{manara13}
{Manara}, C.~F., {Testi}, L., {Rigliaco}, E., {et~al.} 2013,
  \href{http://dx.doi.org/10.1051/0004-6361/201220921}{\JournalTitle{\aap},
  551, A107}

\bibitem[{{Marleau} \& {Aoyama}(2022)}]{ma22}
{Marleau}, G.-D., \& {Aoyama}, Y. 2022,
  \href{http://dx.doi.org/10.3847/2515-5172/acaa34}{\JournalTitle{RNAAS}, 6,
  262}

\bibitem[{{Marleau} {et~al.}(2023){Marleau}, {Kuiper}, {B{\'e}thune}, \&
  {Mordasini}}]{m22Schock}
{Marleau}, G.-D., {Kuiper}, R., {B{\'e}thune}, W., \& {Mordasini}, C. 2023,
  \href{http://dx.doi.org/10.3847/1538-4357/accf12}{\JournalTitle{\apj}, 952,
  89}

\bibitem[{{Marleau} {et~al.}(2022){Marleau}, {Aoyama}, {Kuiper}, {Follette},
  {Turner}, {Cugno}, {Manara}, {Haffert}, {Kitzmann}, {Ringqvist}, {Wagner},
  {van Boekel}, {Sallum}, {Janson}, {Schmidt}, {Venuti}, {Lovis}, \&
  {Mordasini}}]{maea21}
{Marleau}, G.-D., {Aoyama}, Y., {Kuiper}, R., {et~al.} 2022,
  \href{http://dx.doi.org/10.1051/0004-6361/202037494}{\JournalTitle{\aap},
  657, A38}

\bibitem[{{Marois} {et~al.}(2008){Marois}, {Macintosh}, {Barman}, {Zuckerman},
  {Song}, {Patience}, {Lafreni{\`e}re}, \& {Doyon}}]{marois08}
{Marois}, C., {Macintosh}, B., {Barman}, T., {et~al.} 2008,
  \href{http://dx.doi.org/10.1126/science.1166585}{\JournalTitle{Science}, 322,
  1348}

\bibitem[{{Marois} {et~al.}(2010){Marois}, {Zuckerman}, {Konopacky},
  {Macintosh}, \& {Barman}}]{marois10}
{Marois}, C., {Zuckerman}, B., {Konopacky}, Q.~M., {Macintosh}, B., \&
  {Barman}, T. 2010,
  \href{http://dx.doi.org/10.1038/nature09684}{\JournalTitle{\nat}, 468, 1080}

\bibitem[{{Mohanty} {et~al.}(2005){Mohanty}, {Jayawardhana}, \&
  {Basri}}]{mohanty05}
{Mohanty}, S., {Jayawardhana}, R., \& {Basri}, G. 2005,
  \href{http://dx.doi.org/10.1086/429794}{\JournalTitle{\apj}, 626, 498}

\bibitem[{{Mohanty} {et~al.}(2007){Mohanty}, {Jayawardhana}, {Hu{\'e}lamo}, \&
  {Mamajek}}]{mohanty07}
{Mohanty}, S., {Jayawardhana}, R., {Hu{\'e}lamo}, N., \& {Mamajek}, E. 2007,
  \href{http://dx.doi.org/10.1086/510877}{\JournalTitle{\apj}, 657, 1064}

\bibitem[{{Mohanty} {et~al.}(2013){Mohanty}, {Greaves}, {Mortlock}, {Pascucci},
  {Scholz}, {Thompson}, {Apai}, {Lodato}, \& {Looper}}]{mohanty13}
{Mohanty}, S., {Greaves}, J., {Mortlock}, D., {et~al.} 2013,
  \href{http://dx.doi.org/10.1088/0004-637X/773/2/168}{\JournalTitle{\apj},
  773, 168}

\bibitem[{{Mu{\~n}oz} {et~al.}(2020){Mu{\~n}oz}, {Lai}, {Kratter}, \&
  {Miranda}}]{mu20}
{Mu{\~n}oz}, D.~J., {Lai}, D., {Kratter}, K., \& {Miranda}, R. 2020,
  \href{http://dx.doi.org/10.3847/1538-4357/ab5d33}{\JournalTitle{\apj}, 889,
  114}

\bibitem[{{Natta} {et~al.}(2004){Natta}, {Testi}, {Muzerolle}, {Randich},
  {Comer{\'o}n}, \& {Persi}}]{natta04}
{Natta}, A., {Testi}, L., {Muzerolle}, J., {et~al.} 2004,
  \href{http://dx.doi.org/10.1051/0004-6361:20040356}{\JournalTitle{\aap}, 424,
  603}

\bibitem[{{Reggiani} {et~al.}(2016){Reggiani}, {Meyer}, {Chauvin}, {Vigan},
  {Quanz}, {Biller}, {Bonavita}, {Desidera}, {Delorme}, {Hagelberg}, {Maire},
  {Boccaletti}, {Beuzit}, {Buenzli}, {Carson}, {Covino}, {Feldt}, {Girard},
  {Gratton}, {Henning}, {Kasper}, {Lagrange}, {Mesa}, {Messina}, {Montagnier},
  {Mordasini}, {Mouillet}, {Schlieder}, {Segransan}, {Thalmann}, \&
  {Zurlo}}]{reggiani16}
{Reggiani}, M., {Meyer}, M.~R., {Chauvin}, G., {et~al.} 2016,
  \href{http://dx.doi.org/10.1051/0004-6361/201525930}{\JournalTitle{\aap},
  586, A147}

\bibitem[{{Ricci} {et~al.}(2017){Ricci}, {Cazzoletti}, {Czekala}, {Andrews},
  {Wilner}, {Sz{\H{u}}cs}, {Lodato}, {Testi}, {Pascucci}, {Mohanty}, {Apai},
  {Carpenter}, \& {Bowler}}]{ricci17}
{Ricci}, L., {Cazzoletti}, P., {Czekala}, I., {et~al.} 2017,
  \href{http://dx.doi.org/10.3847/1538-3881/aa78a0}{\JournalTitle{\aj}, 154,
  24}

\bibitem[{{Ringqvist} {et~al.}(2023){Ringqvist}, {Viswanath}, {Aoyama},
  {Janson}, {Marleau}, \& {Brandeker}}]{ringqvist23}
{Ringqvist}, S.~C., {Viswanath}, G., {Aoyama}, Y., {et~al.} 2023,
  \href{http://dx.doi.org/10.1051/0004-6361/202245424}{\JournalTitle{\aap},
  669, L12}

\bibitem[{{Rodriguez} {et~al.}(2015){Rodriguez}, {van der Plas}, {Kastner},
  {Schneider}, {Faherty}, {Mardones}, {Mohanty}, \& {Principe}}]{rodriguez15}
{Rodriguez}, D.~R., {van der Plas}, G., {Kastner}, J.~H., {et~al.} 2015,
  \href{http://dx.doi.org/10.1051/0004-6361/201527031}{\JournalTitle{\aap},
  582, L5}

\bibitem[{{Salyk} {et~al.}(2013){Salyk}, {Herczeg}, {Brown}, {Blake},
  {Pontoppidan}, \& {van Dishoeck}}]{salyk13}
{Salyk}, C., {Herczeg}, G.~J., {Brown}, J.~M., {et~al.} 2013,
  \href{http://dx.doi.org/10.1088/0004-637X/769/1/21}{\JournalTitle{\apj}, 769,
  21}

\bibitem[{{Skemer} {et~al.}(2011){Skemer}, {Close}, {Sz{\H{u}}cs}, {Apai},
  {Pascucci}, \& {Biller}}]{skemer11}
{Skemer}, A.~J., {Close}, L.~M., {Sz{\H{u}}cs}, L., {et~al.} 2011,
  \href{http://dx.doi.org/10.1088/0004-637X/732/2/107}{\JournalTitle{\apj},
  732, 107}

\bibitem[{{Tanigawa} {et~al.}(2012){Tanigawa}, {Ohtsuki}, \&
  {Machida}}]{tanigawa12}
{Tanigawa}, T., {Ohtsuki}, K., \& {Machida}, M.~N. 2012,
  \href{http://dx.doi.org/10.1088/0004-637X/747/1/47}{\JournalTitle{\apj}, 747,
  47}

\bibitem[{{Thanathibodee} {et~al.}(2019){Thanathibodee}, {Calvet}, {Bae},
  {Muzerolle}, \& {Hern{\'a}ndez}}]{thanathibodee19}
{Thanathibodee}, T., {Calvet}, N., {Bae}, J., {Muzerolle}, J., \&
  {Hern{\'a}ndez}, R.~F. 2019,
  \href{http://dx.doi.org/10.3847/1538-4357/ab44c1}{\JournalTitle{\apj}, 885,
  94}

\bibitem[{{Thanathibodee} {et~al.}(2022){Thanathibodee}, {Calvet},
  {Hern{\'a}ndez}, {Mauc{\'o}}, \& {Brice{\~n}o}}]{thanathibodee22}
{Thanathibodee}, T., {Calvet}, N., {Hern{\'a}ndez}, J., {Mauc{\'o}}, K., \&
  {Brice{\~n}o}, C. 2022,
  \href{http://dx.doi.org/10.3847/1538-3881/ac3ee6}{\JournalTitle{\aj}, 163,
  74}

\bibitem[{{Theissen} {et~al.}(2018){Theissen}, {Burgasser}, {Bardalez
  Gagliuffi}, {Hardegree-Ullman}, {Gagn{\'e}}, {Schmidt}, \&
  {West}}]{theissen18}
{Theissen}, C.~A., {Burgasser}, A.~J., {Bardalez Gagliuffi}, D.~C., {et~al.}
  2018, \href{http://dx.doi.org/10.3847/1538-4357/aaa0cf}{\JournalTitle{\apj},
  853, 75}

\bibitem[{{Theissen} {et~al.}(2017){Theissen}, {West}, {Shippee}, {Burgasser},
  \& {Schmidt}}]{theissen17}
{Theissen}, C.~A., {West}, A.~A., {Shippee}, G., {Burgasser}, A.~J., \&
  {Schmidt}, S.~J. 2017,
  \href{http://dx.doi.org/10.3847/1538-3881/153/3/92}{\JournalTitle{\aj}, 153,
  92}

\bibitem[{{Venuti} {et~al.}(2019){Venuti}, {Stelzer}, {Alcal{\'a}}, {Manara},
  {Frasca}, {Jayawardhana}, {Antoniucci}, {Argiroffi}, {Natta}, {Nisini},
  {Randich}, \& {Scholz}}]{venuti19}
{Venuti}, L., {Stelzer}, B., {Alcal{\'a}}, J.~M., {et~al.} 2019,
  \href{http://dx.doi.org/10.1051/0004-6361/201935745}{\JournalTitle{\aap},
  632, A46}

\bibitem[{{Wagner} {et~al.}(2018){Wagner}, {Follete}, {Close}, {Apai}, {Gibbs},
  {Keppler}, {M{\"u}ller}, {Henning}, {Kasper}, {Wu}, {Long}, {Males},
  {Morzinski}, \& {McClure}}]{wagner18}
{Wagner}, K., {Follete}, K.~B., {Close}, L.~M., {et~al.} 2018,
  \href{http://dx.doi.org/10.3847/2041-8213/aad695}{\JournalTitle{\apjl}, 863,
  L8}

\bibitem[{{Wagner} {et~al.}(2023){Wagner}, {Stone}, {Skemer}, {Ertel}, {Dong},
  {Apai}, {Spalding}, {Leisenring}, {Sitko}, {Kratter}, {Barman}, {Marley},
  {Miles}, {Boccaletti}, {Assani}, {Bayyari}, {Uyama}, {Woodward}, {Hinz},
  {Briesemeister}, {Lawson}, {M{\'e}nard}, {Pantin}, {Russell}, {Skrutskie}, \&
  {Wisniewski}}]{wagner23}
{Wagner}, K., {Stone}, J., {Skemer}, A., {et~al.} 2023,
  \href{http://dx.doi.org/10.1038/s41550-023-02028-3}{\JournalTitle{Nature
  Astronomy}, 7, 1208}

\bibitem[{{Wiese} \& {Fuhr}(2009)}]{Wiese+Fuhr2009}
{Wiese}, W.~L., \& {Fuhr}, J.~R. 2009,
  \href{http://dx.doi.org/10.1063/1.3077727}{\JournalTitle{Journal of Physical
  and Chemical Reference Data}, 38, 565}

\end{thebibliography}
\bibliographystyle{yahapj.bst}

\end{document}